\documentstyle[11pt,epsfig]{article}
\def\cite#1{#1}
\newcommand{\ct}[1]{[\cite{#1}]}
\def\thebibliography#1{\section*{References}\list
 {[\arabic{enumi}]}{\settowidth\labelwidth{[#1]}\leftmargin\labelwidth
 \advance\leftmargin\labelsep
 \usecounter{enumi}}
 \def\newblock{\hskip .11em plus .33em minus -.07em}
 \sloppy
 \sfcode`\.=1000\relax}

\title {Nucleon electromagnetic structure revisited}

\author{Stanislav~Dubni\v{c}ka$^1${\footnote{e-mail: fyzidubn@savba.sk}}, Anna Zuzana~Dubni\v{c}kov\'a$^2${\footnote{e-mail: dubnickova@fmph.uniba.sk}},
 \\ and  Peter Weisenpacher$^3${\footnote{e-mail: upsyweis@savba.sk}}}
\date{\empty}

\begin{document}
\maketitle
\begin{center} {
$^{1}$ \it Inst. of Physics, Slovak Academy of Sciences,  Bratislava,
Slovak Republic \\
$^{2}$ \it Dept. of Theor. Physics, Comenius Univ., Bratislava,
Slovak Republic \\
$^{3}$ \it Inst. of Informatics, Slovak Academy of Sciences, Bratislava,
Slovak Republic }\\
\end{center}

\vspace{0.5cm}

\def\sa{\omega}
\def\sbb{{\omega^{'}}}
\def\sc{{\omega^{''}}}
\def\sd{{\phi}}
\def\se{{\phi^{'}}}
\def\va{{\varrho}}
\def\vb{{\varrho^{'}}}
\def\vc{{\varrho^{''}}}
\def\vd{{\varrho^{'''}}}
\def\ve{{\varrho^{''''}}}

\def\sam{\omega_0}
\def\sbbm{{\omega^{'}_0}}
\def\scm{{\omega^{''}_0}}
\def\sdm{{\phi_0}}
\def\sem{{\phi^{'}_0}}
\def\vam{{\varrho_0}}
\def\vbm{{\varrho^{'}_0}}
\def\vcm{{\varrho^{''}_0}}
\def\vdm{{\varrho^{'''}_0}}
\def\vem{{\varrho^{''''}_0}}

\def\mmm#1{\frac{m_{#1}^2}{m_{#1}^2-t}}
\def\mmt#1#2{\frac{m_{#1}^2m_{#2}^2}{(m_{#1}^2-t)(m_{#2}^2-t)}}
\def\mtt#1#2#3{\frac{m_{#1}^2m_{#2}^2m_{#3}^2}{(m_{#1}^2-t)(m_{#2}^2-t)
     (m_{#3}^2-t)}}
\def\mc#1{{m^2_#1}}
\def\zll#1#2#3#4{{\frac{#1-#2}{#3-#4}}}
\def\zlll#1#2#3#4#5#6#7#8{{\frac{(#1-#2)(#3-#4)}{(#5-#6)(#7-#8)}}}
\def\zzl#1#2#3#4#5#6{{\frac{#1#2}{(#3-#4)(#5-#6)}}}
\def\xxx#1#2{\frac{(#1_N-#1_#2)(#1_N-#1_#2^*)(#1_N-1/#1_#2)(#1_N-1/#1_#2^*)}
               {(#1-#1_#2)(#1-#1_#2^*)(#1-1/#1_#2)(#1-1/#1_#2^*)}}
\def\eee#1#2{\frac{(#1_N-#1_#2)(#1_N-#1_#2^*)(#1_N+#1_#2)(#1_N+#1_#2^*)}
               {(#1-#1_#2)(#1-#1_#2^*)(#1+#1_#2)(#1+#1_#2^*)}}
\def\cc#1{{C_#1}}
\def\zl#1#2#3#4#5{{\frac{#2^#1-#3^#1}{#4^#1-#5^#1}}}
\def\uvv#1{{(\frac{1-#1^2}{1-#1_N^2})}}
\def\fff#1#2{(f^{(#1)}_{#2{NN}}/f_{#2})}
\def\ccc#1#2#3#4#5{\frac{C^{#1}_#2-C^{#1}_#3}{C^{#1}_#4-C^{#1}_#5}}

\def\mcsa{{\mc\sa}}
\def\mcsb{{\mc\sbb}}
\def\mcsd{{\mc\sd}}
\def\mcva{{\mc\va}}
\def\mcvb{{\mc\vb}}
\def\mcvc{{\mc\vc}}
\def\mcsc{{\mc\sc}}
\def\mcse{{\mc\se}}
\def\mcvd{{\mc\vd}}
\def\mcve{{\mc\ve}}

\def\ccsa{{\cc\sa}}
\def\ccsb{{\cc\sbb}}
\def\ccsd{{\cc\sd}}
\def\ccva{{\cc\va}}
\def\ccvb{{\cc\vb}}
\def\ccvc{{\cc\vc}}
\def\ccsc{{\cc\sc}}
\def\ccse{{\cc\se}}
\def\ccvd{{\cc\vd}}
\def\ccve{{\cc\ve}}

\begin{abstract}

   Unitary and analytic ten-resonance model of the nucleon electromagnetic
(e.m.) structure with canonical normalizations and QCD (up to the logarithmic
correction) asymptotics is constructed on the four-sheeted Riemann surface,
which provides a superposition of vector-meson pole and continuum contributions
in a very natural way. As a result it describes simultaneously all existing experimental
space-like and time-like data on the proton e.m. form factors (ff's) and on the neutron e.m.
ff's as well. A crucial factor in the latter achievement is the inclusion of a contribution of
the fourth excited state of the $\rho(770)$ meson with the parameters $m_{\rho''''}=2455\pm53 MeV,
\Gamma_{\rho''''}=728\pm2 MeV$ and $(f^{(1)}_{\rho''''NN}/f_{\rho''''})=0.0549\pm0.0005$,
$(f^{(2)}_{\rho''''NN}/f_{\rho''''})=-0.0103\pm0.0001$.
The pronounced effect of the two-pion continuum on the isovector spectral functions
demonstrating a strong enhancement of the left wing of the $\rho(770)$-resonance close to
two-pion threshold, which was revealed by H\"ohler and Pietarinen by means of the nucleon ff
unitarity condition more than a quarter of the century ago, is predicted by the model automatically.
The model gives large values of the $f^{(1,2)}_{\phi NN}$ coupling constants, thus
indicating the violation of the OZI rule. Since in the framework of the considered
model isoscalar ff's above their lowest branch point  $t^s_0=9m^2_{\pi}$ are complex functions,
the isoscalar spectral function behaviours are predicted as well.
\end{abstract}

\newpage

\section{Introduction}

    The electromagnetic (e.m.) structure of the nucleons, as revealed first time in
elastic electron-nucleon scattering almost half a century ago, is completely
described by four independent scalar functions of one variable  called form factors (ff's).
They depend on the square momentum transfer $t=-Q^2$ of the virtual photon.

   Nucleon e.m. ff's can be chosen in a divers way, e.g. as the Dirac and Pauli
ff's, $F^p_1(t)$, $F^n_1(t)$ and $F^p_2(t)$, $F^n_2(t)$, or the Sachs
electric and magnetic ff's, $G^p_E(t)$, $G^n_E(t)$ and $G_M^p(t)$, $G^n_M(t)$,
or isoscalar and isovector Dirac and Pauli ff's, $F_1^s(t)$, $F_1^v(t)$ and
$F_2^s(t)$, $F^v_2(t)$ and isoscalar and isovector electric and magnetic ff's,
$G_E^s(t)$, $G_E^v(t)$ and $G^s_M(t)$, $G^v_M(t)$, respectively.

   The Dirac and Pauli ff's are naturally obtained in a decomposition of the
nucleon matrix element of the e.m. current into a maximum number of linearly
independent covariants constructed from the four-momenta, $\gamma$-matrices
and Dirac bispinors of nucleons as follows

\begin{equation}
\langle N|J^{e.m.}_\mu|N\rangle=e\bar u(p')\{\gamma_\mu
F^N_1(t)+\frac{i}{2m_N}\sigma_{\mu\nu}(p'-p)_{\nu}F^N_2(t)\}u(p)\label{d1}
\end{equation}
with $m_N$ to be  nucleon mass.

   On the other hand, the electric and magnetic ff's are very suitable in
extracting  experimental information on the nucleon e.m. structure
from the measured cross sections

\begin{equation}
\frac{d\sigma^{lab}(e^-N\to
e^-N)}{d\Omega}=\frac{\alpha^2}{4E^2}\frac{\cos^2(\theta/2)}{\sin^4(\theta/2)}
\frac{1}{1+(\frac{2E}{m_N})\sin^2(\theta/2)}[\frac{G^2_E-\frac{t}{4m_N^2}G^2_M}
{1-\frac{t}{4m_N^2}}-2\frac{t}{4m_N^2}G^2_M\tan^2(\theta/2)]
\label{d2}
\end{equation}

$\alpha=1/137$, $E$-the incident electron energy,\\
and

\begin{equation}
\sigma_{tot}^{c.m.}(e^+e^-\to N\bar N)=\frac{4\pi\alpha^2\beta_N}{3t}
[|G_M(t)|^2+\frac{2m_N^2}{t}|G_E(t)|^2],\;\;\; \beta_N=\sqrt{1-\frac{4m_N^2}{t}}
\label{d3}
\end{equation}
or

\begin{equation}
\sigma_{tot}^{c.m.}(\bar p p\to e^+e^-)=\frac{2\pi\alpha^2}{3p_{c.m.}\sqrt{t}}
[|G_M(t)|^2+\frac{2m_N^2}{t}|G_E(t)|^2],
\label{d4}
\end{equation}
($p_{c.m.}$-antiproton momentum in the c.m. system)\\
as there are no interference terms between them.

   In the Breit frame, the Sachs ff's give the distribution of charge and
magnetization within the proton and neutron, respectively. From all
four Sachs ff's the neutron electric ff plays a particular role. Though
the total neutron charge is zero, there is a nonvanishing distribution of
charge, which leads to the nonvanishing neutron electric ff.

   The isoscalar and isovector Dirac and Pauli ff's are suitable for
a construction of various phenomenological models of the nucleon e.m.
structure. The most attractive  of them is the Vector Meson Dominance (VMD)
picture  in the framework of which ff's are simply saturated by a set of
isoscalar and isovector vector meson poles on the positive real axis. However,
this turns out to be practically an insufficient approximation and in a more
realistic description of the data (especially in the time-like region)  instability of
vector-mesons has to be taken into account and the contributions of continua, to be created by
n-particle thresholds, like, e.g., $2\pi, 3\pi, K \bar K, N \bar N$ etc., together with
the correct asymptotic behaviours and normalizations have to be included.

   In recent years, abundant and very accurate data on the nucleon e.m. ff's
appeared. Most of the references concerning the nucleon space-like data
can be found in \ct{1}. More recent precise measurements are presented in
\ct{2-10}. Besides the latter, there are also very accurate data on the ratio
of proton electric and magnetic ff's \ct{11} obtained at the Jefferson Lab for
$-0.5 GeV^2 < -3.5 GeV^2$ by using polarization transfer and the new data
on the neutron electric ff from BATES \ct{12}, MAMI \ct{13-16} and NIKHEF
\ct{17}.

   For the time-like region data see \ct{18-26}. There, in particular,
the FENICE experiment in Frascati measured, besides the proton e.m. ff's
\ct{25}, the magnetic neutron ff in the time-like region \ct{26} for
the first time.
 There are also valuable results on the proton magnetic ff
    at higher energies measured at FERMILAB \ct{21,22}.

   All this  stimulated  recent dispersion theoretical analysis \ct{27,28} of the
nucleon e.m. ff data in the space-like region and in the time-like region
\ct{29} as well. The latter works are an update and extension of historically
the most competent nucleon ff analysis carried out by H\"ohler with
collaborators \ct{30}. However, the model does not allow one to describe all the
time-like data consistently, while still giving  good description of the
data in the space-like region.

   In this paper, we construct a ten-resonance unitary and analytic model
of the nucleon e.m. structure, defined on the four-sheeted Riemann surface with canonical
normalizations and QCD asymptotics, which provides a very effective framework for
a superposition of complex conjugate vector-meson pole pairs on unphysical sheets and
continua contributions in nucleon e.m. ff's . The model contains, e.g., an explicit two-pion
continuum contribution given by the unitary cut starting with $t=4m_\pi^2$ and automatically
predicts the strong enhancement of the left wing of the $\rho(770)$ resonance
in the isovector spectral functions to be consistent with the results of
\ct{27,31}.

   Another result of the presented model is the prediction of parameters of the fourth
excited state of the  $\rho(770)$ meson and the automatic prediction of
isoscalar nucleon spectral function behaviours. At the same time, a description
of all existing space-like and time-like nucleon e.m. ff data, including also FENICE
(Frascati) results on the neutron from the $e^+e^-\to n\bar n$ process, is achieved.

   The paper is organized as follows. In  Section 2., the unitary and analytic
ten-resonance model of the nucleon e.m. structure with canonical
normalizations and asymptotics as predicted by the quark model of
hadrons is constructed. Evaluation of all free parameters of the
model (however, with clear physical meaning) by a fit of all
existing data is carried out in  Section 3. In Section 4., we
predict the isoscalar and isovector nucleon spectral function
behaviours and various coupling constant ratio values, which
appear not to be free parameters of the constructed model. The
last section is devoted to conclusions and discussion.

\section{Ten-resonance unitary and analytic model of nucleon e.m. structure}

    All four independent sets of four nucleon e.m. ff's discussed in the
Introduction are related by

\begin{eqnarray}
\nonumber G_E^p(t)&=&G_E^s(t)+G_E^v(t)=F_1^p(t)+\frac{t}{4m^2_p}F_2^p(t)=
[F_1^s(t)]+ F_1^v(t)]+\frac{t}{4m_p^2}[F_2^s(t)+F_2^v(t)];\\
 G_M^p(t)&=&G_M^s(t)+G_M^v(t)=F_1^p(t)+F_2^p(t)=[F_1^s(t)+F_1^v(t)]+
[F_2^s(t)+F_2^v(t)]; \label{d5} \\
\nonumber G_E^n(t)&=&G_E^s(t)-G_E^v(t)=F_1^n(t)+\frac{t}{4m_n^2}F_2^n(t)=
[F_1^s(t)-F_1^v(t)]+\frac{t}{4m_n^2}[F_2^s(t)-F_2^v(t)];\\
\nonumber G_M^n(t)&=&G^s_M(t)-G_M^v(t)=F_1^n(t)+F_2^n(t)=[F_1^s(t)-F_1^v(t)]+
[F_2^s(t)-F_2^v(t)],
\end{eqnarray}
and at the value $t=0$ normalized as follows:

\begin{eqnarray}
(i)&\nonumber &
\ G_E^p(0)=1;\ G_M^p(0)=1+\mu_p;\ G_E^n(0)=0;\ G_M^n(0)=\mu_n;\\
(ii)&\nonumber &
\ G_E^s(0)=G_E^v(0)=\frac{1}{2};\ G_M^s(0)=\frac{1}{2}(1+\mu_p+\mu_n);\
 G_M^v(0)=\frac{1}{2}(1+\mu_p-\mu_n);\\
(iii)&\label{d6} &
\ F_1^p(0)=1;\ F_2^p(0)=\mu_p;\ F_1^n(0)=0;\ F_2^n(0)=\mu_n;\\
(iv)&\nonumber&\ F_1^s(0)=F_1^v(0)=\frac{1}{2};\
 F_2^s(0)=\frac{1}{2}(\mu_p+\mu_n);\
 F_2^v(0)=\frac{1}{2}(\mu_p-\mu_n),
\end{eqnarray}
where $\mu_p$ and $\mu_n$ are the proton and neutron anomalous magnetic
moments, respectively.

   The ten-resonance unitary and analytic model will represent a consistent
unification of the following three fundamental features (besides other
properties) of the nucleon e.m. ff's:
\begin{itemize}
\item[1.]
The experimental fact of  creation of unstable vector-meson resonances
in the $e^+e^-$-annihilation processes into hadrons.
\item[2.]
The hypothetical analytic properties of the nucleon e.m. ff's on the first (physical) sheet
of the Riemann surface, by means of which just the contributions of continua are taken into
account.
\item[3.]
The asymptotic behaviour of nucleon e.m. ff's as predicted \ct{32} by the
quark model of hadrons.
\end{itemize}

   Here we would like to note that a further procedure will not mean any mathematically correct
derivation of the unitary and analytic model, but only an (noncommutative) algorithm of its
construction which is, however, generally valid also for  any other strongly
interacting particles.

   In order to take into account the first feature, one starts with  saturation of the
isoscalar and isovector parts of the Dirac and Pauli ff's by the isoscalar
and isovector vector mesons possessing the quantum numbers of the photon.
As there are no data on the nucleon e.m. ff's for  reliable determination of resonance masses
and widths in the region $0<t<4m_N^2$ of  manifestation of the majority of resonances under
consideration, these parameters are fixed at the world averaged values. Then, their
consistency with existing ff data in other regions is investigated.

   In Review of Particle Physics \ct{33} we find just 5 isoscalar resonances
$\omega$(782), $\phi$(1020), $\omega^{'}$(1420),
$\omega^{''}$(1600), $\phi^{'}$(1680) with the required
properties. However, one finds there only 3 isovector resonances
$\rho$(770), $\rho^{'}$(1450), $\rho^{''}$(1700) with quantum
numbers of the photon. On the other hand, we have gained
experience in our previous analyses that the most stable
description of existing data is achieved if an equal number of
isoscalar and isovector vector meson resonances in the
investigated models is taken into account. Therefore, in the
isovector Dirac and Pauli ff's we consider the third excited state
of the $\rho$-meson, $\rho^{'''}$(2150), revealed in \ct{34}, and
in order to achieve also a description of the time-like region
data on proton at higher energies from Fermilab \ct{21,22} and
neutron data from Frascati \ct{26}, we also introduce
hypothetically the fourth excited state of the $\rho$-meson,
$\rho^{''''}$,  the mass and width of which are left to be free
parameters of the model.

   As one will see later from the comparison of the unitary and analytic model
with all existing data, those resonance parameters will be found to be quite
reasonable, and a simultaneous description of the space-like and time-like
nucleon ff data, including  the FENICE (Frascati) results
on the neutron, will be achieved.

   With the aim of  incorporation of the third feature of nucleon e.m. ff's
we transform VMD parametrizations of the isoscalar and isovector parts of
the Dirac and Pauli ff's into the common denominators. The explicit
requirement of the normalizations (\ref{d6}) and the asymptotic behaviours

\begin{eqnarray}
t^{i+1}F_i^{s,v}(t)_{|t|\to\infty}\sim {constant},\;\;\;\;i=1,2
\label{d7}
\end{eqnarray}
lead (for more detail see Appendix A) again to the zero-width VMD
parametrization of the isoscalar and isovector parts of the Dirac and
Pauli ff's

\begin{eqnarray}
\nonumber F^s_1(t)&=&\frac{1}{2}\mmt\sc\sbb+\\
  \nonumber &+&\left\{\mmt\sc\sa\zll\mcsc\mcsa\mcsc\mcsb-
  \mmt\sbb\sa\zll\mcsb\mcsa\mcsc\mcsb-\right.\\
  &-&\left.\mmt\sc\sbb\right\}\fff 1\sa+\label{d8}\\
  \nonumber &+&\left\{\mmt\sc\sd\zll\mcsc\mcsd\mcsc\mcsb-
  \mmt\sbb\sd\zll\mcsb\mcsd\mcsc\mcsb-\right.\\
  \nonumber &-&\left.\mmt\sc\sbb\right\}\fff 1\sd-\\
  \nonumber &-&\left\{\mmt\se\sc\zll\mcse\mcsc\mcsc\mcsb-
  \mmt\se\sbb\zll\mcse\mcsb\mcsc\mcsb+\right.\\
  \nonumber &+&\left.\mmt\sc\sbb\right\}\fff 1\se,\\
\nonumber & &\qquad\\
\nonumber F^v_1(t)&=&\frac{1}{2}\mmt\vc\vb+\\
   \nonumber &+&\left\{\mmt\vc\va\zll\mcvc\mcva\mcvc\mcvb-
\mmt\vb\va\zll\mcvb\mcva\mcvc\mcvb-\right.\\
   &-&\left.\mmt\vc\vb\right\}\fff 1\va+\label{d9}\\
   \nonumber &+&\left\{\mmt\vd\vb\zll\mcvd\mcvb\mcvc\mcvb-
\mmt\vd\vc\zll\mcvd\mcvc\mcvc\mcvb-\right.\\
   \nonumber &-&\left.\mmt\vc\vb\right\}\fff 1\vd-\\
   \nonumber &-&\left\{\mmt\ve\vc\zll\mcve\mcvc\mcvc\mcvb-
\mmt\ve\vb\zll\mcve\mcvb\mcvc\mcvb+\right.\\
   \nonumber &+&\left.\mmt\vc\vb\right\}\fff 1\ve,\\
\nonumber & &\qquad\\
\nonumber F_2^s(t)&=&\frac{1}{2}(\mu_p+\mu_n)\mtt\sc\sbb\sa+\\
   \nonumber &+&\left\{\mtt\sc\sd\sa\zlll\mcsc\mcsd\mcsd\mcsa\mcsc\mcsb\mcsb\mcsa+\right.\\
   \nonumber &+&\mtt\sc\sbb\sd\zlll\mcsc\mcsd\mcsb\mcsd\mcsc\mcsa\mcsb\mcsa-\\
   \nonumber &-&\mtt\sbb\sd\sa\zlll\mcsb\mcsd\mcsd\mcsa\mcsc\mcsb\mcsc\mcsa-\\
   &-&\left.\mtt\sc\sbb\sa\right\}\fff 2\sd+\label{d10}\\
   \nonumber &+&\left\{\mtt\se\sc\sbb\zlll\mcse\mcsc\mcse\mcsb\mcsc\mcsa\mcsb\mcsa-\right.\\
   \nonumber &-&\mtt\se\sc\sa\zlll\mcse\mcsc\mcse\mcsa\mcsc\mcsb\mcsb\mcsa+\\
   \nonumber &+&\mtt\se\sbb\sa\zlll\mcse\mcsb\mcse\mcsa\mcsc\mcsb\mcsc\mcsa-\\
   \nonumber &-&\left.\mtt\sc\sbb\sa\right\}\fff 2\se,\\
\nonumber & &\qquad\\
\nonumber F_2^v(t)&=&\frac{1}{2}(\mu_p-\mu_n)\mtt\vc\vb\va+\\
  \nonumber &+&\left\{\mtt\vd\vb\va\zlll\mcvd\mcvb\mcvd\mcva\mcvc\mcvb\mcvc\mcva-\right.\\
  \nonumber &-&\mtt\vd\vc\va\zlll\mcvd\mcvc\mcvd\mcva\mcvc\mcvb\mcvb\mcva+\\
  \nonumber &+&\mtt\vd\vc\vb\zlll\mcvd\mcvc\mcvd\mcvb\mcvc\mcva\mcvb\mcva-\\
  &-&\left.\mtt\vc\vb\va\right\}\fff 2\vd+\label{d11}\\
  \nonumber &+&\left\{\mtt\ve\vb\va\zlll\mcve\mcvb\mcve\mcva\mcvc\mcvb\mcvc\mcva-\right.\\
  \nonumber &-&\mtt\ve\vc\va\zlll\mcve\mcvc\mcve\mcva\mcvc\mcvb\mcvb\mcva+\\
  \nonumber &+&\mtt\ve\vc\vb\zlll\mcve\mcvc\mcve\mcvb\mcvc\mcva\mcvb\mcva-\\
  \nonumber &-&\left.\mtt\vc\vb\va\right\}\fff 2\ve.
\end{eqnarray}
However, they are already automaticly normalized and they govern the
asymptotics (7) as predicted by QCD up to the logarithmic corrections.

   Despite the latter properties the model is unable to reproduce the
existing experimental information properly and only its unitarization, i.e.,  inclusion of the
contributions of continua and instability of vector-meson resonances, leads to a simultaneous
description of the space-like and time-like data.

   It is well known that the unitarity condition requires the imaginary part
of the nucleon e.m. ff's to be different from zero only above the lowest branch
point $t_0$ and, moreover, it just predicts its smoothly varying behaviour
(see e.g. \ct{27,31}).

   The unitarization of the model (\ref{d8})-(\ref{d11}) can be
achieved by application of the following special non-linear transformations
\begin{eqnarray}
\nonumber t=t_0^s-\frac{4(t_{in}^{1s}-t_0^s)}{[1/V-V]^2}\\
t=t_0^v-\frac{4(t_{in}^{1v}-t_0^v)}{[1/W-W]^2}\label{d12}\\
\nonumber t=t_0^s-\frac{4(t_{in}^{2s}-t_0^s)}{[1/U-U]^2}\\
\nonumber t=t_0^v-\frac{4(t_{in}^{2v}-t_0^v)}{[1/X-X]^2},\nonumber
\end{eqnarray}
respectively, and a subsequent incorporation of the nonzero values of
vector meson widths.

   Here $t^s_0=9m_{\pi}^2, t^v_0=4m_{\pi}^2, t_{in}^{1s}, t_{in}^{1v}, t_{in}^{2s}, t_{in}^{2v}$
are square-root branch points, as it is transparent from the inverse
transformations to (\ref{d12}), e.g

\begin{equation}
V(t)=i\frac
{\sqrt{\left (\frac{t_{in}^{1s}-t_0^s}{t_0^s}\right )^{1/2}+\left (\frac{t-t_0^s}{t_0^s}\right )^{1/2}}-
 \sqrt{\left (\frac{t_{in}^{1s}-t_0^s}{t_0^s}\right )^{1/2}-\left (\frac{t-t_0^s}{t_0^s}\right )^{1/2}}}
{\sqrt{\left (\frac{t_{in}^{1s}-t_0^s}{t_0^s}\right )^{1/2}+\left (\frac{t-t_0^s}{t_0^s}\right )^{1/2}}+
 \sqrt{\left (\frac{t_{in}^{1s}-t_0^s}{t_0^s}\right )^{1/2}-\left (\frac{t-t_0^s}{t_0^s}\right )^{1/2}}}
\label{d13}
\end{equation}
and similarly for $W(t), U(t)$ and $X(t)$.

   The interpretation of $t^s_0=9m_{\pi}^2$ and $t^v_0=4m_{\pi}^2$ is clear. They are the lowest
branch points of isoscalar and isovector Dirac and Pauli ff's on the positive real axis,
respectively, as in the isoscalar case the 3-pion states and in the isovector case the 2-pion
states are the lowest intermediate mass states in the unitarity conditions of the corresponding
ff's.

   However, as it follows just from the unitarity conditions of ff's, there is an
infinite number of allowed higher mass intermediate states and as a result there is an infinite
number of the corresponding branch points ( and thus, an infinite number of branch cut contributions ) in
every of the considered nucleon ff's.

    Since, in principle, an infinite number of cuts cannot be taken into account in any theoretical
scheme, we restrict ourselves in every isoscalar and isovector Dirac and Pauli ff to the two-cut
approximation. The second one, an effective inelastic cut, in every isoscalar and isovector Dirac
and Pauli ff is generated just by the square-root branch points  $t_{in}^{1s}, t_{in}^{1v},
t_{in}^{2s}, t_{in}^{2v}$, respectively. They are free parameters of the model and the data
themselves, by a fitting procedure, will choose for them such numerical values  that the
contributions of the corresponding square-root cuts will be practically equivalent to
the contributions of an infinite number of unitary branch cuts in every considered ff.

   Some experts are suggesting to fix these square-root branch points at the $N \bar N$
threshold. However, it will be demonstrated in  Section 3 that
this can be done only in the case of isovector parts of Dirac and
Pauli ff's, but  in none of the cases of  the isoscalar parts of
the Dirac and Pauli ff's.

   So, by  application of (\ref{d12}) to (\ref{d8})-(\ref{d11}), for every isoscalar and
isovector Dirac and Pauli ff one gets one analytic function in the whole complex $t$-plane
besides two right-hand cuts (see Appendix B) of the following forms:
\begin{eqnarray}
F^s_1[V(t)]&=&\left(\frac{1-V^2}{1-V_N^2}\right)^4\left\{\frac{1}{2}
H_{\omega''}(V)\cdot L_{\omega'}(V)+ \left [H_{\omega''}(V)\cdot L_{\omega}(V)\cdot
\frac{C^{1s}_{\omega''}-C^{1s}_{\omega}}{C^{1s}_{\omega''}-C^{1s}_{\omega'}}\right.\right. - \nonumber  \\
&-& L_{\omega'}(V)\cdot L_{\omega}(V)\ccc {1s}\sbb\sa\sc\sbb
- H_{\omega''}(V)\cdot L_{\omega'}(V)\biggr ]\fff 1\sa +  \nonumber \\
&+&\left [H_{\omega''}(V)\cdot L_{\phi}(V)\ccc{1s}\sc\sd\sc\sbb -
 L_{\omega'}(V)\cdot L_{\phi}(V)\ccc{1s}\sbb\sd\sc\sbb\right. - \label{d14} \\
&-& H_{\omega''}(V)\cdot L_{\omega'}(V)\biggr ]\fff 1\sd-
\left [H_{\phi'}(V)\cdot H_{\omega''}(V)\ccc{1s}\se\sc\sc\sbb\right. - \nonumber \\
&-& H_{\phi'}(V)\cdot L_{\omega'}(V)\ccc{1s}\se\sbb\sc\sbb +
  H_{\omega''}(V)\cdot L_{\omega'}(V)\biggr ]\fff 1\se\biggr\}  \nonumber
\end{eqnarray}

\begin{eqnarray}
 F^v_1[W(t)]&=&\left(\frac{1-W^2}{1-W_N^2}\right)^4\left\{\frac{1}{2}
L_{\rho''}(W)\cdot L_{\rho'}(W)+ \left [L_{\rho''}(W)\cdot L_{\rho}(W)\ccc{1v}\vc\va\vc\vb -\right.\right. \nonumber \\
&-& L_{\rho'}(W)\cdot L_{\rho}(W)\ccc{1v}\vb\va\vc\vb -
 L_{\rho''}(W)\cdot L_{\rho'}(W)\biggr]\fff 1\va+ \nonumber \\
  &+&\left[H_{\rho'''}(W)\cdot L_{\rho'}(W)\ccc{1v}\vd\vb\vc\vb\right.
 -H_{\rho'''}(W)\cdot L_{\rho''}(W)\ccc{1v}\vd\vc\vc\vb -  \nonumber \\
 &-&L_{\rho''}(W)\cdot L_{\rho'}(W)\biggr ]\fff 1\vd- \label{d15} \\
 &-&\left [H_{\rho''''}(W)\cdot L_{\rho''}(W)\ccc{1v}\ve\vc\vc\vb \right.
 -H_{\rho''''}(W)\cdot L_{\rho'}(W)\ccc{1v}\ve\vb\vc\vb + \nonumber   \\
  &+&L_{\rho''}(W)\cdot L_{\rho'}(W)\biggr]\fff 1\ve\biggr\} \nonumber
\end{eqnarray}

\begin{eqnarray}
 F_2^s[U(t)]&=&\left(\frac{1-U^2}{1-U_N^2}\right)^6\left\{\frac{1}{2}(\mu_p+\mu_n)
H_{\omega''}(U)\cdot L_{\omega'}(U)\cdot L_{\omega}(U)\right.+  \nonumber    \\
 &+&\left[H_{\omega''}(U)\cdot L_{\phi}(U)\cdot L_{\omega}(U)\ccc{2s}\sc\sd\sc\sbb.\ccc {2s}\sd\sa\sbb\sa\right. + \nonumber \\
  &+&H_{\omega''}(U)\cdot L_{\omega'}(U)\cdot L_{\phi}(U) \ccc{2s}\sc\sd\sc\sa.\ccc {2s}\sbb\sd\sbb\sa- \nonumber  \\
&-&L_{\omega'}(U)\cdot L_{\phi}(U)\cdot L_{\omega}(U)\ccc{2s}\sbb\sd\sc\sbb.\ccc {2s}\sd\sa\sc\sa-   \nonumber \\
 &-&H_{\omega''}(U)\cdot L_{\omega'}(U)\cdot L_{\omega}(U)\biggr]\fff 2\sd+
\label{d16} \\
 &+&\left[H_{\phi'}(U)\cdot H_{\omega''}(U)\cdot
L_{\omega'}(U)\ccc{2s}\se\sc\sc\sa.\ccc {2s}\se\sbb\sbb\sa \right.- \nonumber   \\
 &-&H_{\phi'}(U)\cdot H_{\omega''}(U)\cdot L_{\omega}(U)\ccc{2s}\se\sc\sc\sbb.\ccc {2s}\se\sa\sbb\sa+  \nonumber   \\
  &+&H_{\phi'}(U)\cdot L_{\omega'}(U)\cdot L_{\omega}(U)\ccc{2s}\se\sbb\sc\sbb.\ccc {2s}\se\sa\sc\sa- \nonumber \\
&-&H_{\omega''}(U)\cdot L_{\omega'}(U)\cdot L_{\omega}(U)\biggr]\fff 2\se\biggr\} \nonumber
\end{eqnarray}

\begin{eqnarray}
F_2^v[X(t)]&=&\left(\frac{1-X^2}{1-X_N^2}\right)^6\left\{\frac{1}{2}(\mu_p-\mu_n)
L_{\rho''}(X)\cdot L_{\rho'}(X)\cdot L_{\rho}(X)\right. +  \nonumber    \\
&+&\left [H_{\rho'''}(X)\cdot L_{\rho'}(X)\cdot L_{\rho}(X)\ccc{2v}\vd\vb\vc\vb.\ccc {2v}\vd\va\vc\va\right.-  \nonumber  \\
&-&H_{\rho'''}(X)\cdot L_{\rho''}(X)\cdot L_{\rho}(X)\ccc{2v}\vd\vc\vc\vb.\ccc {2v}\vd\va\vb\va+   \nonumber   \\
 &+&H_{\rho'''}(X)\cdot L_{\rho''}(X)\cdot L_{\rho'}(X)\ccc{2v}\vd\vc\vc\va.\ccc {2v}\vd\vb\vb\va-  \nonumber     \\
  &-&L_{\rho''}(X)\cdot L_{\rho'}(X)\cdot L_{\rho}(X) \biggr ]\fff 2\vd+
\label{d17}    \\
 &+&\left[H_{\rho''''}(X)\cdot L_{\rho'}(X)\cdot L_{\rho}(X)\ccc{2v}\ve\vb\vc\vb.\ccc {2v}\ve\va\vc\va\right.-  \nonumber     \\
  &-&H_{\rho''''}(X)\cdot L_{\rho''}(X)\cdot L_{\rho}(X)\ccc{2v}\ve\vc\vc\vb.\ccc {2v}\ve\va\vb\va+ \nonumber  \\
 &+&H_{\rho''''}(X)\cdot L_{\rho''}(X)\cdot L_{\rho'}(X)\ccc{2v}\ve\vc\vc\va.\ccc {2v}\ve\vb\vb\va- \nonumber    \\
 &-&L_{\rho''}(X)\cdot L_{\rho'}(X)\cdot L_{\rho}(X)\biggr ]\fff 2\ve\biggr\} \nonumber
\end{eqnarray}
where
\begin{eqnarray}
& &L_r(V)=\frac{(V_N-V_r)(V_N-V_r^*)(V_N-1/V_r)(V_N-1/V_r^*)}
               {(V-V_r)(V-V_r^*)(V-1/V_r)(V-1/V_r^*)}; \label{d18} \\
\nonumber & &C_r^{1s}=\frac{(V_N-V_r)(V_N-V_r^*)(V_N-1/V_r)(V_N-1/V_r^*)}
               {-(V-1/V_r)(V-1/V_r^*)};\qquad r=\omega,\phi,\omega^{'},\\
& &H_l(V)=\frac{(V_N-V_l)(V_N-V_l^*)(V_N+V_l)(V_N+V_l^*)}
               {(V-V_l)(V-V_l^*)(V+V_l)(V+V_l^*)}; \label{d19} \\
\nonumber & &C_l^{1s}=\frac{(V_N-V_l)(V_N-V_l^*)(V_N+V_l)(V_N+V_l^*)}
               {-(V_l-1/V_l)(V_l^*-1/V_l^*)};\qquad l=\omega^{''},\phi^{'} \\
& & L_k(W)=\frac{(W_N-W_k)(W_N-W_k^*)(W_N-1/W_k)(W_N-1/W_k^*)}
               {(W-W_k)(W-W_k^*)(W-1/W_k)(W-1/W_k^*)}; \label{d20} \\
\nonumber & &C_k^{1v}=\frac{(W_N-W_k)(W_N-W_k^*)(W_N-1/W_k)(W_N-1/W_k^*)}
               {-(W_k-1/W_k)(W_k^*-1/W_k^*)}; \qquad k=\rho,\rho^{'},\rho^{''},\\
& & H_n(W)=\frac{(W_N-W_n)(W_N-W_n^*)(W_N+W_n)(W_N+W_n^*)}
               {(W-W_n)(W-W_n^*)(W+W_n)(W+W_n^*)}; \label{d21} \\
\nonumber & &C_n^{1v}=\frac{(W_N-W_n)(W_N-W_n^*)(W_N+W_n)(W_N+W_n^*)}
               {-(W_n-1/W_n)(W_n^*-1/W_n^*)}; \qquad n=\rho^{'''},\rho^{''''}\nonumber \\
& &L_r(U)=\frac{(U_N-U_r)(U_N-U_r^*)(U_N-1/U_r)(U_N-1/U_r^*)}
               {(U-U_r)(U-U_r^*)(U-1/U_r)(U-1/U_r^*)}; \label{d22} \\
\nonumber & &C_r^{2s}=\frac{(U_N-U_r)(U_N-U_r^*)(U_N-1/U_r)(U_N-1/U_r^*)}
               {-(U-1/U_r)(U-1/U_r^*)};\qquad r=\omega,\phi,\omega^{'},\\
& &H_l(U)=\frac{(U_N-U_l)(U_N-U_l^*)(U_N+U_l)(U_N+U_l^*)}
               {(U-U_l)(U-U_l^*)(U+U_l)(U+U_l^*)}; \label{d23} \\
\nonumber & &C_l^{2s}=\frac{(U_N-U_l)(U_N-U_l^*)(U_N+U_l)(U_N+U_l^*)}
               {-(U_l-1/U_l)(U_l^*-1/U_l^*)};\qquad l=\omega^{''},\phi^{'} \\
& & L_k(X)=\frac{(X_N-X_k)(X_N-X_k^*)(X_N-1/X_k)(X_N-1/X_k^*)}
               {(X-X_k)(X-X_k^*)(X-1/X_k)(X-1/X_k^*)}; \label{d24} \\
\nonumber & &C_k^{2v}=\frac{(X_N-X_k)(X_N-X_k^*)(X_N-1/X_k)(X_N-1/X_k^*)}
               {-(X_k-1/X_k)(X_k^*-1/X_k^*)}; \qquad k=\rho,\rho^{'},\rho^{''},\\
& & H_n(X)=\frac{(X_N-X_n)(X_N-X_n^*)(X_N+X_n)(X_N+X_n^*)}
               {(X-X_n)(X-X_n^*)(X+X_n)(X+X_n^*)}; \label{d25} \\
\nonumber & &C_n^{2v}=\frac{(X_N-X_n)(X_N-X_n^*)(X_N+X_n)(X_N+X_n^*)}
               {-(X_n-1/X_n)(X_n^*-1/X_n^*)}; \qquad n=\rho^{'''},\rho^{''''} \\
\nonumber
\end{eqnarray}

Expressions (\ref{d14})-(\ref{d17}), together with  relations
(\ref{d5}), represent just the ten-resonance unitary and analytic
model of the nucleon e.m. structure with canonical normalizations
(\ref{d6}) and the correct asymptotic behaviours as predicted by
the quark model of hadrons. In the next section this model is used
to analyze  all existing nucleon e.m. ff data and  obtain further
predictions.

\section{Analysis of all existing space-like and time-like data}

   Taking into account the discussion in Section 2 and applying the asymptotic
conditions \ct{35} together with ff normalizations, the ten-resonance unitary and analytic model
of the nucleon e.m. structure depends (see Appendix A) on the following parameters:

\begin{eqnarray}
t_{in}^{1s}, t_{in}^{1v}, t_{in}^{2s}, t_{in}^{2v},
m_{\rho^{''''}}, \Gamma_{\rho^{''''}},
(f_{\omega NN}^{(1)}/f_{\omega}), (f_{\phi NN}^{(1)}/f_{\phi}),
(f_{\phi^{'} NN}^{(1)}/f_{\phi^{'}}), (f_{\rho NN}^{(1)}/f_{\rho}),\label{d26} \\
(f_{\rho^{'''} NN}^{(1)}/f_{\rho^{'''}}),
(f_{\rho^{''''} NN}^{(1)}/f_{\rho^{''''}}),
(f_{\phi NN}^{(2)}/f_{\phi}), (f_{\phi^{'} NN}^{(2)}/f_{\phi^{'}}),
(f_{\rho^{'''} NN}^{(2)}/f_{\rho^{'''}}),
(f_{\rho^{''''} NN}^{(2)}/f_{\rho^{''''}}).\nonumber
\end{eqnarray}

\begin{figure}[thb]
\begin{center}
\psfig{figure=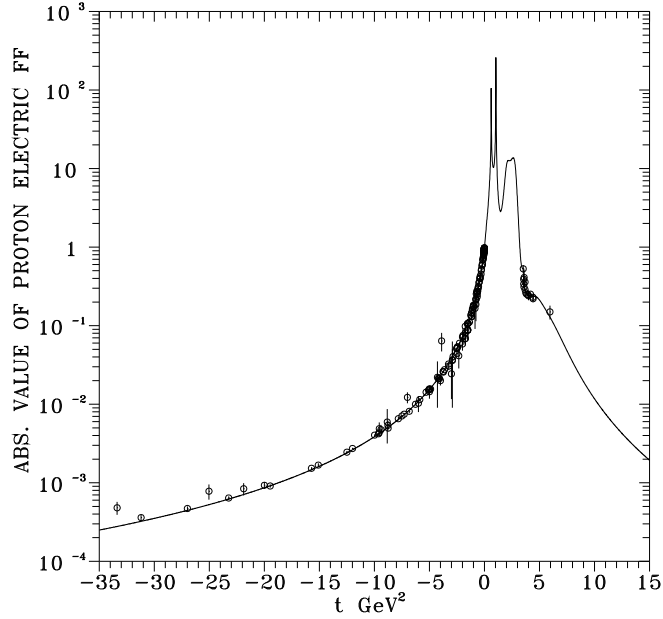,width=9cm}
\end{center}
\caption{ A simultanious optimal fit of all existing data on
proton electric ff}
%ff's in the space-like and time-like regions by the unitary and analytic ten resonance
%model of the proton e.m. structure, represented by expressions (5) and
%(30)-(33).}
%\label{bariic}
\end{figure}
\begin{figure}[thb]
\begin{center}
\psfig{figure=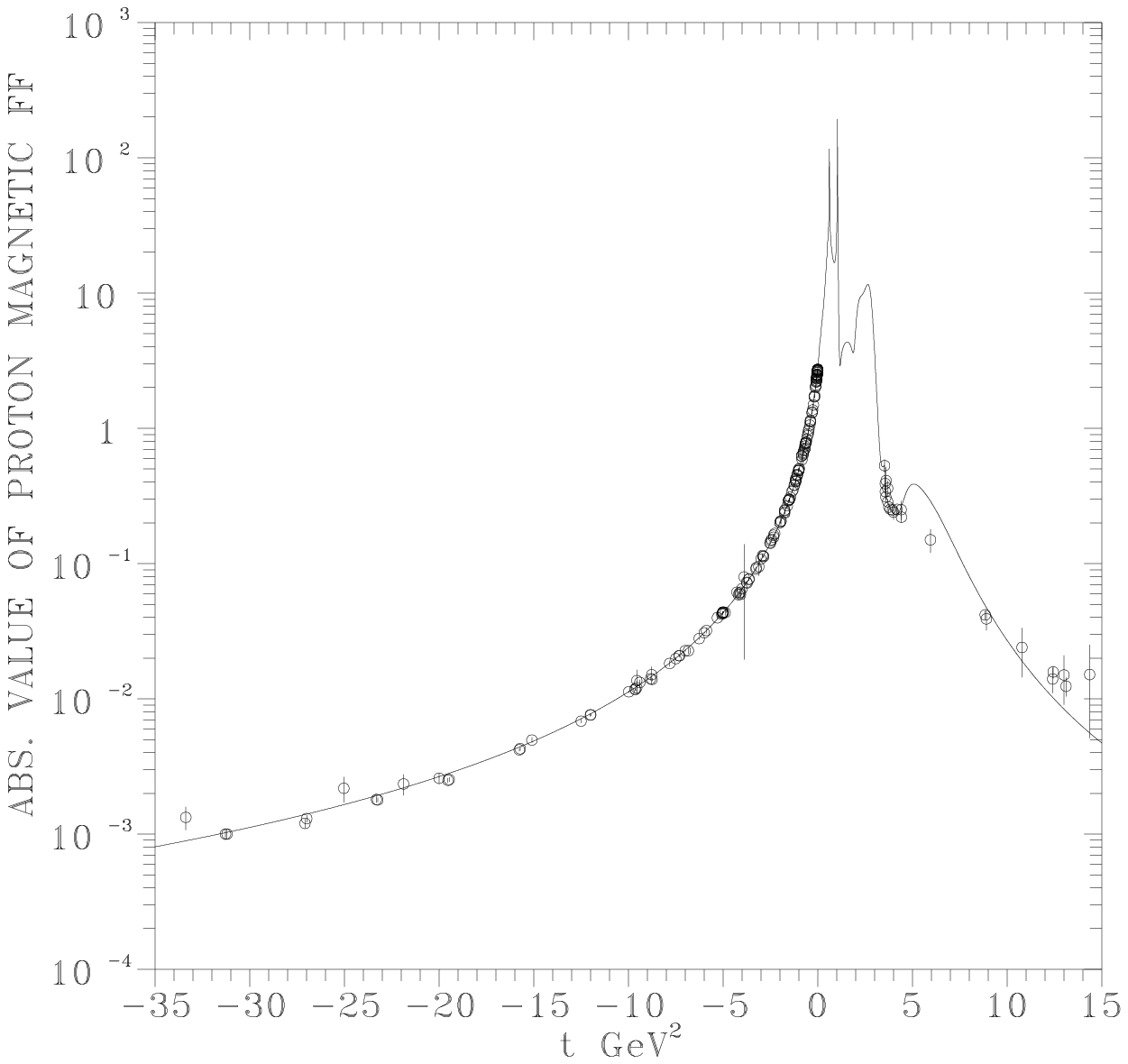,width=9cm}
\end{center} \caption{A simultanious optimal fit of all existing
data on proton magnetic ff}
%ff's in the space-like and time-like regions by the unitary and analytic ten resonance
%model of the proton e.m. structure, represented by expressions (5) and
%(30)-(33).}
%\label{bariic}
\end{figure}

   Here we would like to note that  criticism has been expressed by  some
specialists that the presented unitary and analytic model of nucleon e.m. structure comprises
too many (though with clear physical meaning) free parameters. We believe that the responsibility
for such a situation rests with various experimental groups confirming \ct{33} nowadays so many
vector-meson resonances possessing the quantum numbers of the photon, as a substantial majority
of free parameters of the model are coupling constats just of these vector-mesons under
consideration with nucleons. Once these vector-meson resonances exist, they can not be ignored
in any theoretical considerations, including theoretical models of the nucleon e.m. structure.
\begin{figure}[thb]
\begin{center}
\psfig{figure=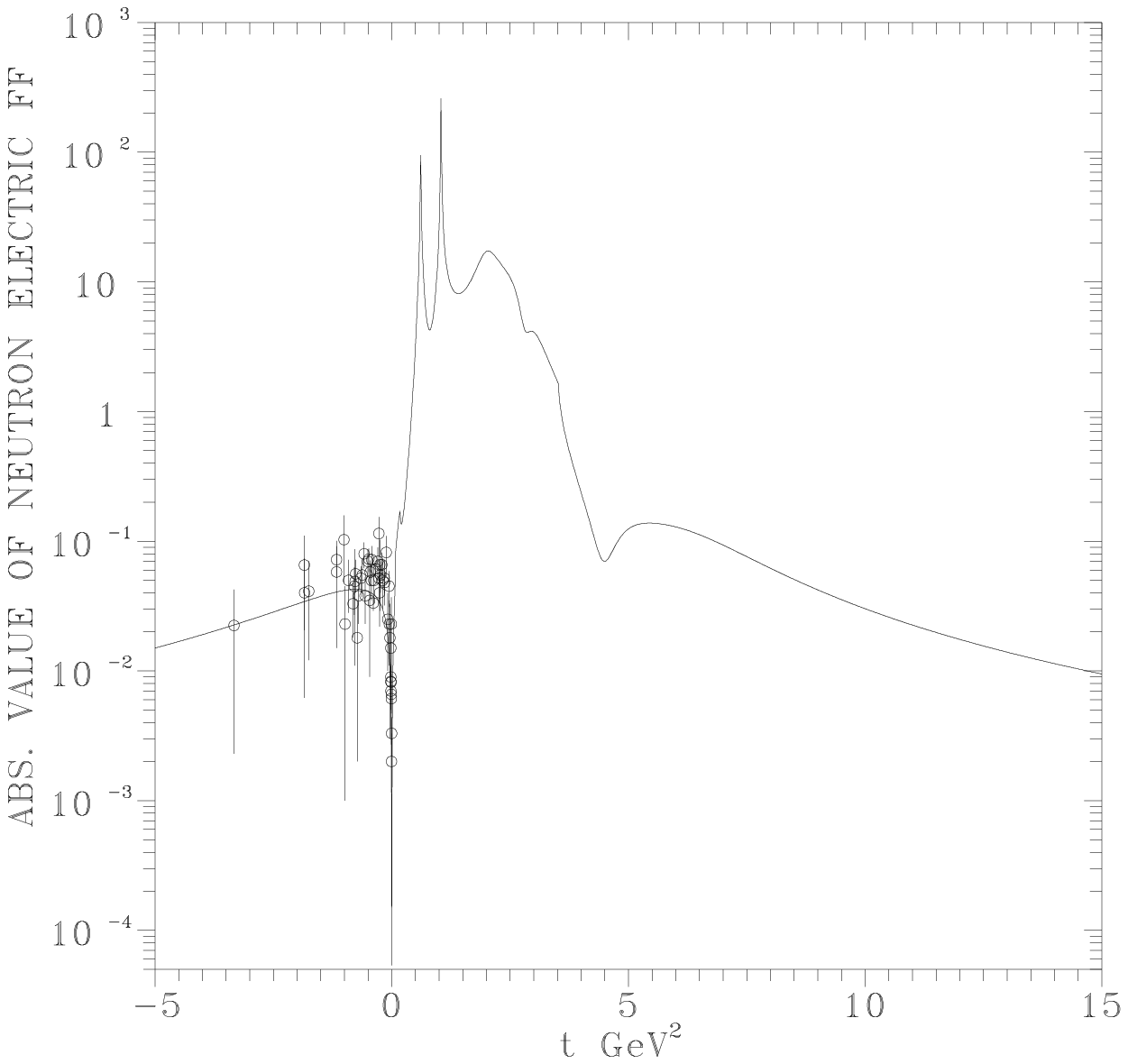,width=9cm}
\end{center} \caption{A simultanious optimal fit of all existing
data on neutron electric ff}
%ff's in the space-like and time-like regions by the unitary and analytic ten resonance
%model of the proton e.m. structure, represented by expressions (5) and
%(30)-(33).}
%\label{bariic}
\end{figure}

\begin{figure}[thb]
\begin{center}
\psfig{figure=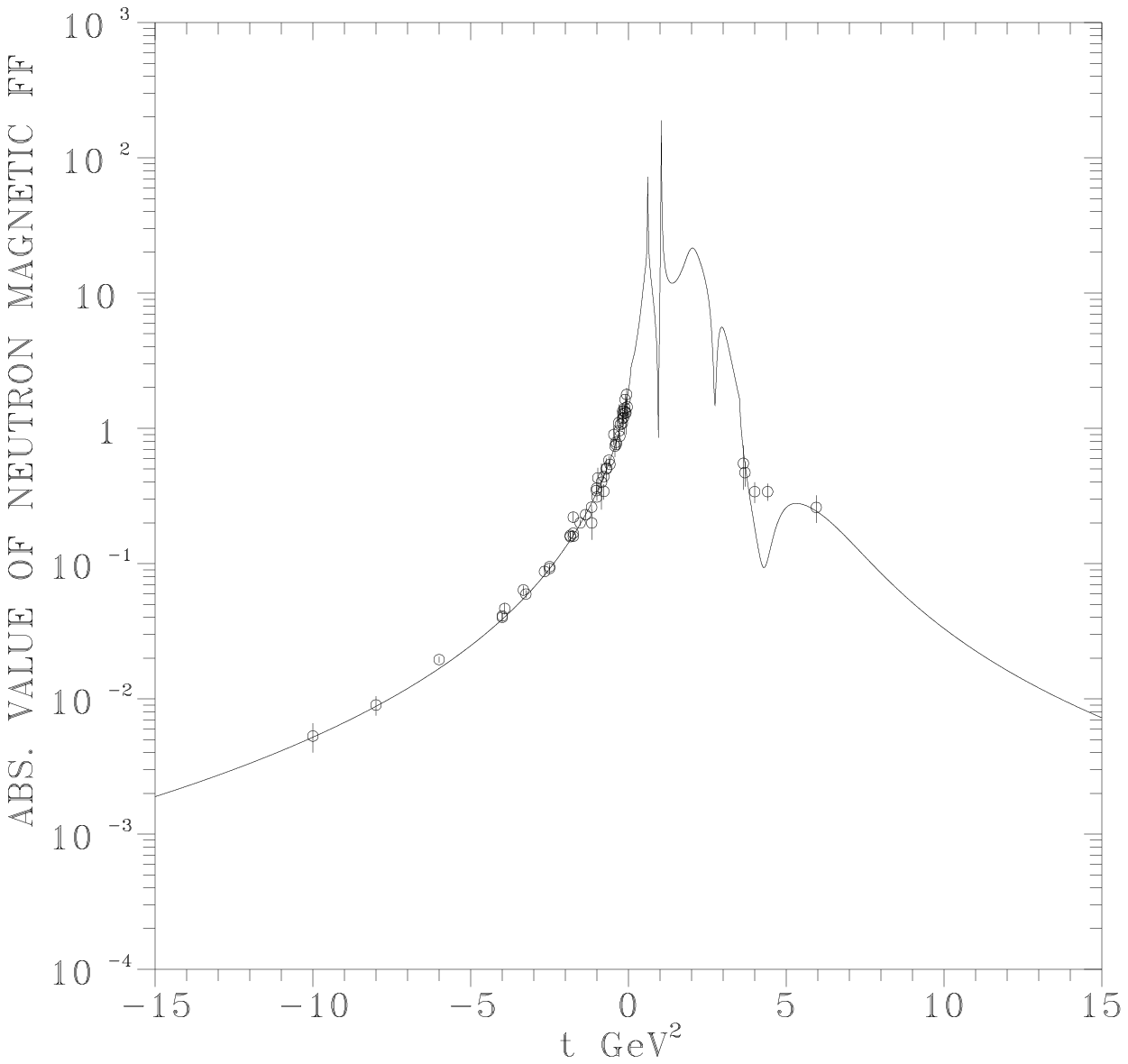,width=9cm}
\end{center} \caption{A simultanious optimal fit of all existing
data on neutron magnetic ff}
%ff's in the space-like and time-like regions by the unitary and analytic ten resonance
%model of the proton e.m. structure, represented by expressions (5) and
%(30)-(33).}
%\label{bariic}
\end{figure}

   A solution for  lowering  the number of free parameters could be true numerical
values of these coupling constants, however, brought from outside of the considered model.
Nevertheless, for the determination of coupling constants there is no reliable theory up to now
and as a consequence, one is forced to consider them to be free parameters of the model.

   For numerical evaluation of the parameters (\ref{d26}) we have collected 512 experimental
points  which have been discussed in more detail in the Introduction. The data have been analyzed
by   relations (\ref{d5}) and (\ref{d14})-(\ref{d17}) by using the CERN program
MINUIT. The best description of them was achieved with $\chi^2/ndf=1.46$ and
the following values of free parameters:
\begin{eqnarray}
\nonumber
&t_{in}^{1s}=2.6012\pm 0.6391\;GeV^2 &t_{in}^{1v}=3.5220\pm 0.0059\;GeV^2\\
\nonumber
&t_{in}^{2s}=2.7200\pm 0.6271\;GeV^2 &t_{in}^{2v}=3.6316\pm 0.6235\;GeV^2\\
\nonumber
&(f_{\omega NN}^{(1)}/f_{\omega})=1.1112\pm 0.0030 &(f_{\rho NN}^{(1)}/f_{\rho})=0.3843\pm 0.0043\\
\nonumber
&(f_{\phi NN}^{(1)}/f_{\phi})=-0.9389\pm 0.0056 &(f_{\rho^{'''} NN}^{(1)}/f_{\rho^{'''}})=-0.0840\pm 0.0008\\
&(f_{\phi^{'} NN}^{(1)}/f_{\phi^{'}})=-0.3255\pm 0.0047 &(f_{\rho^{'''} NN}^{(2)}/f_{\rho^{'''}})=0.0299\pm 0.0003
\label{d27}\\
\nonumber
&(f_{\phi NN}^{(2)}/f_{\phi})=-0.2659\pm 0.0287 &m_{\rho^{''''}}=2455\pm 53\;MeV\\
\nonumber
&(f_{\phi^{'} NN}^{(2)}/f_{\phi^{'}})=0.1190\pm 0.0032 &\Gamma_\ve=728\pm 2\;MeV\\
\nonumber
&\;&(f_{\rho^{''''} NN}^{(1)}/f_{\rho^{''''}})=0.0549\pm 0.0005\\
\nonumber
&\;&(f_{\rho^{''''} NN}^{(2)}/f_{\rho^{''''}})=-0.0103\pm 0.0001.
\end{eqnarray}

   If, on the basis of suggestions of some experts, the parameters
$t_{in}^{1s}, t_{in}^{1v}, t_{in}^{2s}, t_{in}^{2v}$ are fixed at the $N \bar N$ threshold,
the best description of existing data is achieved with a worse value of $\chi^2/ndf=1.82$
and the rest parameters as follows:
\begin{eqnarray}
\nonumber
&(f_{\omega NN}^{(1)}/f_{\omega})=0.9916\pm 0.0112 &(f_{\rho NN}^{(1)}/f_{\rho})=0.3746\pm 0.0159\\
\nonumber
&(f_{\phi NN}^{(1)}/f_{\phi})=-1.1209\pm 0.0125 &
(f_{\rho^{'''} NN}^{(1)}/f_{\rho^{'''}})=-0.0799\pm 0.0006\\
&(f_{\phi^{'} NN}^{(1)}/f_{\phi^{'}})=-2.6079\pm 0.0384 &
(f_{\rho^{'''} NN}^{(2)}/f_{\rho^{'''}})=0.0324\pm 0.0002
\label{d28}\\
\nonumber
&(f_{\phi NN}^{(2)}/f_{\phi})=-4.4532\pm 0.1020 & m_{\rho^{''''}}=2461\pm 38\;MeV\\
\nonumber
&(f_{\phi^{'} NN}^{(2)}/f_{\phi^{'}})=0.3617\pm 0.0502 &\Gamma_\ve=728\pm 44\;MeV\\
\nonumber
&\;&(f_{\rho^{''''} NN}^{(1)}/f_{\rho^{''''}})=0.0542\pm 0.0004\\
\nonumber
&\;&(f_{\rho^{''''} NN}^{(2)}/f_{\rho^{''''}})=-0.0112\pm 0.0001.
\end{eqnarray}

   From the comparison of the numerical values of (\ref{d28}) with (\ref{d27}) we
come to the conclusions  that by  fixing the second branch points of the isoscalar and
isovector Dirac and Pauli ff's at the nucleon-antinucleon threshold the coupling constant
ratios of isovector vector-mesons to nucleons (the mass and the width of $\rho{''''}$ as well) are
almost unchanged, but the coupling constant ratios of isoscalar vector-mesons to nucleons
(especially of higher vector-mesons) are remarkably out of order.

   This fact has  explanation in the values of parameters of (\ref{d27})  where the thresholds
$t_{in}^{1s}, t_{in}^{1v}, t_{in}^{2s}, t_{in}^{2v}$ were left to be found in a fitting procedure
of data. In the isovector case, they were determined around the value corresponding to the
nucleon-antinucleon threshold and, as a result, it does not matter if they are fixed at the
$N \bar N$ threshold or they are found almost at the same position in a fitting procedure.
However, in the isoscalar case they are found at  much lower values than the $N \bar N$
threshold. So, those values indicate (unlike the isovector ff's) that between the
lowest $t_0^s=9m_{\pi}^2$ branch point and the $N \bar N$ threshold there is some
allowed intermediate mass state in the unitarity condition generating an important
cut contribution  which cannot be neglected in a description of the nucleon e.m. structure.
We know from other considerations that it is just the $K \bar K$ threshold.

    Compilation of world nucleon ff data and their description by our
ten-resonance unitary and analytic model with parameters (\ref{d27}) is graphically represented
in Figs. 1-4. One can see from Fig. 4 that unlike   papers \ct{27,29}
the ten-resonance unitary and analytic model is able to describe FENICE time-like data on
neutron \ct{26} quite well. The same is also valid  for the FERMILAB proton time-like data
\ct{21,22} (see Fig. 2). The latter was possible to achieve by  introducing  a hypothetical
fourth excited state of the $\rho$(770)-meson the parameters of which were
found in a fitting procedure of all existing data to be quite reasonable (see (\ref{d27})).
Its existence, however, has to be proved by being identified
also in other processes and not only $e^+e^-\to N\bar N$.

   Of particular interest is  determination of the electric and magnetic, and also Dirac and
Pauli radii of nucleons. They are given in Table 1  where for  comparison the results of  papers
\ct{27} and \ct{30} are presented too.

\begin{table}[thb]
\caption{\bf Electric and magnetic, and also Dirac and Pauli radii
of the proton and neutron}
\begin{center}
%\begin{minipage}{12cm}
\begin{tabular}{|c|c|c|c|c|c|c|} \hline
&$r^{p}_E[fm]$ & $r^{p}_M[fm]$ & $r^{n}_M[fm]$ & $r^{p}_1[fm]$ &
$r^{p}_2[fm]$ & $r^{n}_2[fm]$\\ \hline \hline our
results&0.827&0.860&0.891&0.752&0.914&0.883 \\ \hline Ref.
[27]&0.847&0.836&0.889&0.774&0.894&0.893 \\ \hline Ref.
[30]&0.836&0.843&0.840&0.761&0.883&0.876\\ \hline
\end{tabular}
%\end{minipage}
\end{center}
\end{table}

   Here we would like to stress that the neutron charge radius is predicted by the model to be
negative automatically and its value $<r^2_{E,n}>=-0.130 [fm^2]$ is more or less compatible with
the newest experimental result $<r^2_{E,n}>=-0.113\pm 0.003\pm 0.004 [fm^2]$ \ct{36}.

   In order to demonstrate explicitly substantial deviations from the dipole
fit in all channels and at the same time the violation of the
nucleon ff scaling, particularly at large momentum transfer, we
show in Figs. 5 the electric and magnetic proton and neutron ff's
in the space-like region normalized to the dipole formula
$G_D(t)=(1-t/0.71)^{-2}$.

\begin{figure}[thb]
\begin{center}
\psfig{figure=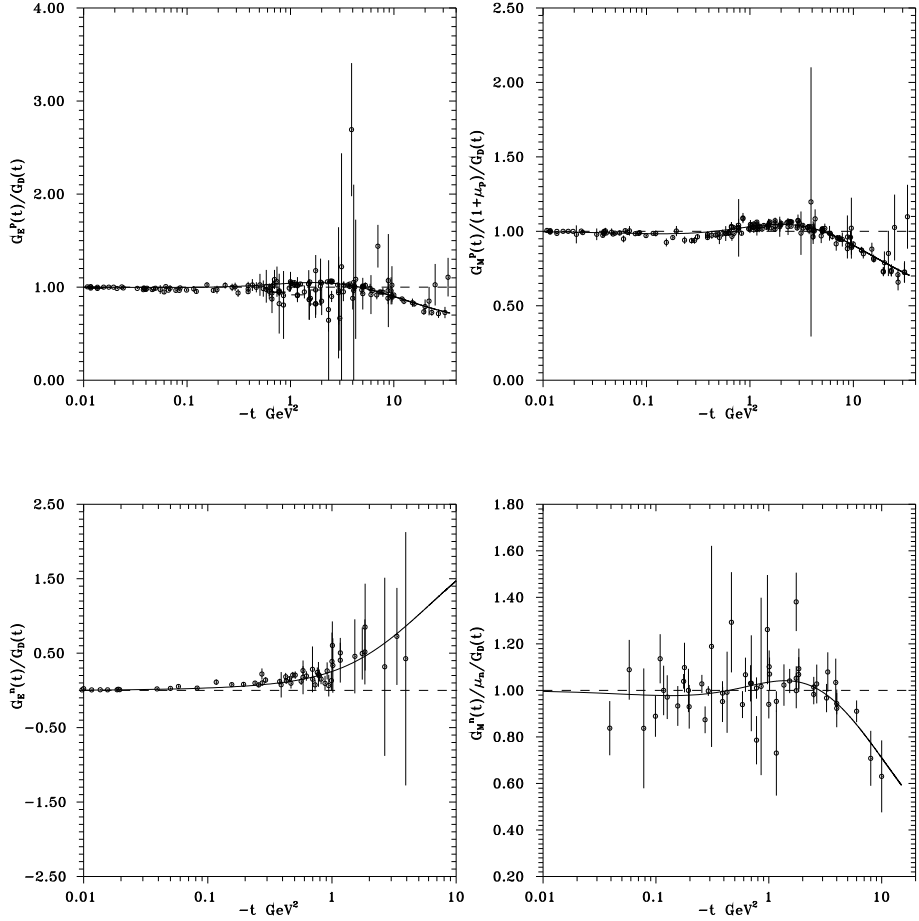,width=13cm}
\end{center} \caption{ Ratios of appropriately normalized electric
and magnetic proton and neutron ff's in the space-like region to
the dipole formula.}
%\label{bar}
\end{figure}

   In relation to the data on decreasing ratio of proton electric and magnetic ff's \ct{11}
obtained recently at the Jefferson Lab for $-3.5 GeV^2 <t< -0.5 GeV^2$ by using polarization
transfer, one can say that the unitary and analytic ten-resonance model at the same values of
momentum transfers gives almost negligible falling around the value 1.02 . However, the latter
result does not mean that the model presented here is in  disagreement with the Jefferson Lab
data, but all experimental points measured up to present time   (see Figs. 1 and 2) in the
interval $-3.5 GeV^2 <t< -0.5 GeV^2$ are in  contradiction with them. If the data
on the electric and magnetic proton ff in the space-like region are consistent with the new
Jefferson Lab data, then the  behaviours of $G_E^p(t)$ and $G_M^p(t)$ predicted by the unitary
and analytic model will be consistent with them too.

\section{Other predictions of unitary and analytic model of the nucleon e.m. structure}

   The unitary and analytic ten-resonance model of the nucleon e.m. structure
constructed in this paper reflects all known nucleon ff properties and thus, it gives one
analytic function, i.e. one smooth function on the whole real axis for every nucleon e.m. ff.
As a result, one can then believe  the predicted behaviours of these nucleon e.m. ff's to be
realistic also outside the regions of existing experimental data.

   Valuable is the predicted existence of the fourth excited state of
the $\rho(770)$-meson with the resonance parameters $m_{\rho^{''''}}=2455\pm 38\;MeV$
and $\Gamma_{\rho^{''''}}=728\pm 2\;MeV$ without  which one could not achieve
a satisfactory description of the FENICE time-like neutron data \ct{26} and
also of eight FERMILAB proton experimental points \ct{21,22} at higher energies.

   Taking into account the numerical results (\ref{d27}) for the parameters of the model
and the transformed relations for additional coupling constant ratios (\ref{a6})-(\ref{a9})
from Appendix A

\begin{eqnarray}
\nonumber {\rm I}.\;\;\; (f^{(1)}_{\omega^{'} NN}/f_{\omega^{'}})&=&\frac{1}{2}
 \frac{C^{1s}_{\omega^{''}}}{C^{1s}_{\omega^{''}}-C^{1s}_{\omega^{'}}}
 -(f^{(1)}_{\omega^{} NN}/f_{\omega^{}})
 \frac{C^{1s}_{\omega^{''}}-C^{1s}_{\omega^{}}}
 {C^{1s}_{\omega^{''}}-C^{1s}_{\omega^{'}}}-\\
&-&(f^{(1)}_{\phi^{} NN}/f_{\phi^{}})
 \frac{C^{1s}_{\omega^{''}}-C^{1s}_{\phi^{}}}
 {C^{1s}_{\omega^{''}}-C^{1s}_{\omega^{'}}}
 +(f^{(1)}_{\phi^{'} NN}/f_{\phi^{'}})
 \frac{C^{1s}_{\phi^{'}}-C^{1s}_{\omega^{'}}}
 {C^{1s}_{\omega^{''}}-C^{1s}_{\omega^{'}}}\label{d29}\\
\nonumber (f^{(1)}_{\omega^{''} NN}/f_{\omega^{''}})&=&-\frac{1}{2}
 \frac{C^{1s}_{\omega^{'}}}{C^{1s}_{\omega^{''}}-C^{1s}_{\omega^{'}}}
 +(f^{(1)}_{\omega^{} NN}/f_{\omega^{}})
 \frac{C^{1s}_{\omega^{'}}-C^{1s}_{\omega^{}}}
 {C^{1s}_{\omega^{''}}-C^{1s}_{\omega^{'}}}+\\
\nonumber &+&(f^{(1)}_{\phi^{} NN}/f_{\phi^{}})
 \frac{C^{1s}_{\omega^{'}}-C^{1s}_{\phi^{}}}
 {C^{1s}_{\omega^{''}}-C^{1s}_{\omega^{'}}}
 -(f^{(1)}_{\phi^{'} NN}/f_{\phi^{'}})
 \frac{C^{1s}_{\phi^{'}}-C^{1s}_{\omega^{'}}}
 {C^{1s}_{\omega^{''}}-C^{1s}_{\omega^{'}}}\\
\nonumber &\ &\ \\
\nonumber {\rm II}.\;\;\; (f^{(1)}_{\rho^{'} NN}/f_{\rho^{'}})&=&\frac{1}{2}
 \frac{C^{1v}_{\rho^{''}}}{C^{1v}_{\rho^{''}}-C^{1v}_{\rho^{'}}}
 -(f^{(1)}_{\rho^{} NN}/f_{\rho^{}})
 \frac{C^{1v}_{\rho^{''}}-C^{1v}_{\rho^{}}}
 {C^{1v}_{\rho^{''}}-C^{1v}_{\rho^{'}}}+\\
&+&(f^{(1)}_{\rho^{'''} NN}/f_{\rho^{'''}})
 \frac{C^{1v}_{\rho^{'''}}-C^{1v}_{\rho^{''}}}
 {C^{1v}_{\rho^{''}}-C^{1v}_{\rho^{'}}}
 +(f^{(1)}_{\rho^{''''} NN}/f_{\rho^{''''}})
 \frac{C^{1v}_{\rho^{''''}}-C^{1v}_{\rho^{''}}}
 {C^{1v}_{\rho^{''}}-C^{1v}_{\rho^{'}}}\label{d30}\\
\nonumber (f^{(1)}_{\rho^{''} NN}/f_{\rho^{''}})&=&-\frac{1}{2}
 \frac{C^{1v}_{\rho^{'}}}{C^{1v}_{\rho^{''}}-C^{1v}_{\rho^{'}}}
 +(f^{(1)}_{\rho^{} NN}/f_{\rho^{}})
 \frac{C^{1v}_{\rho^{'}}-C^{1v}_{\rho^{}}}
 {C^{1v}_{\rho^{''}}-C^{1v}_{\rho^{'}}}-\\
\nonumber &-&(f^{(1)}_{\rho^{'''} NN}/f_{\rho^{'''}})
 \frac{C^{1v}_{\rho^{'''}}-C^{1v}_{\rho^{'}}}
 {C^{1v}_{\rho^{''}}-C^{1v}_{\rho^{'}}}
 -(f^{(1)}_{\rho^{''''} NN}/f_{\rho^{''''}})
 \frac{C^{1v}_{\rho^{''''}}-C^{1v}_{\rho^{'}}}
 {C^{1v}_{\rho^{''}}-C^{1v}_{\rho^{'}}}\\
\nonumber &\ &\ \\
\nonumber {\rm III}.\;\;\; (f^{(2)}_{\omega^{} NN}/f_{\omega^{}})&=&\frac{1}{2}
 (\mu_p+\mu_n)\frac{C^{2s}_{\omega^{''}}C^{2s}_{\omega^{'}}}
 {(C^{2s}_{\omega^{''}}-C^{2s}_{\omega^{}})
  (C^{2s}_{\omega^{'}}-C^{2s}_{\omega^{}})}-\\
\nonumber &-&(f^{(2)}_{\phi^{} NN}/f_{\phi^{}})
 \frac{(C^{2s}_{\omega^{''}}-C^{2s}_{\phi^{}})
       (C^{2s}_{\omega^{'}}-C^{2s}_{\phi^{}})}
       {(C^{2s}_{\omega^{''}}-C^{2s}_{\omega^{}})
       (C^{2s}_{\omega^{'}}-C^{2s}_{\omega^{}})}-\\
\nonumber &-&(f^{(2)}_{\phi^{'} NN}/f_{\phi^{'}})
   \frac{(C^{2s}_{\phi^{'}}-C^{2s}_{\omega^{''}})
       (C^{2s}_{\phi^{'}}-C^{2s}_{\omega^{'}})}
       {(C^{2s}_{\omega^{''}}-C^{2s}_{\omega^{}})
       (C^{2s}_{\omega^{'}}-C^{2s}_{\omega^{}})}\\
\nonumber (f^{(2)}_{\omega^{'} NN}/f_{\omega^{'}})&=&-\frac{1}{2}
 (\mu_p+\mu_n)\frac{C^{2s}_{\omega^{''}}C^{2s}_{\omega^{}}}
 {(C^{2s}_{\omega^{''}}-C^{2s}_{\omega^{'}})
  (C^{2s}_{\omega^{'}}-C^{2s}_{\omega^{}})}-\\
&-&(f^{(2)}_{\phi^{} NN}/f_{\phi^{}})
 \frac{(C^{2s}_{\omega^{''}}-C^{2s}_{\phi^{}})
       (C^{2s}_{\phi^{}}-C^{2s}_{\omega^{}})}
       {(C^{2s}_{\omega^{''}}-C^{2s}_{\omega^{'}})
       (C^{2s}_{\omega^{'}}-C^{2s}_{\omega^{}})}+\label{d31}\\
\nonumber
       &+&(f^{(2)}_{\phi^{'} NN}/f_{\phi^{'}})
  \frac{(C^{2s}_{\phi^{'}}-C^{2s}_{\omega^{''}})
       (C^{2s}_{\phi^{'}}-C^{2s}_{\omega^{}})}
       {(C^{2s}_{\omega^{''}}-C^{2s}_{\omega^{'}})
       (C^{2s}_{\omega^{'}}-C^{2s}_{\omega^{}})}\\
\nonumber (f^{(2)}_{\omega^{''} NN}/f_{\omega^{''}})&=&\frac{1}{2}
 (\mu_p+\mu_n)\frac{C^{2s}_{\omega^{'}}C^{2s}_{\omega^{}}}
 {(C^{2s}_{\omega^{''}}-C^{2s}_{\omega^{'}})
  (C^{2s}_{\omega^{''}}-C^{2s}_{\omega^{}})}+\\
\nonumber
 &+&(f^{(2)}_{\phi^{} NN}/f_{\phi^{}})
       \frac{(C^{2s}_{\omega^{'}}-C^{2s}_{\phi^{}})
       (C^{2s}_{\phi^{}}-C^{2s}_{\omega^{}})}
       {(C^{2s}_{\omega^{''}}-C^{2s}_{\omega^{'}})
       (C^{2s}_{\omega^{''}}-C^{2s}_{\omega^{}})}-\\
\nonumber &-&  (f^{(2)}_{\phi^{'} NN}/f_{\phi^{'}})
  \frac{(C^{2s}_{\phi^{'}}-C^{2s}_{\omega^{'}})
       (C^{2s}_{\phi^{'}}-C^{2s}_{\omega^{}})}
       {(C^{2s}_{\omega^{''}}-C^{2s}_{\omega^{'}})
       (C^{2s}_{\omega^{''}}-C^{2s}_{\omega^{}})}\\
\nonumber &\ &\ \\
{\rm IV}.\;\;\; (f^{(2)}_{\rho^{} NN}/f_{\rho^{}})&=&\frac{1}{2}
 (\mu_p-\mu_n)\frac{C^{2v}_{\rho^{''}}C^{2v}_{\rho^{'}}}
 {(C^{2v}_{\rho^{''}}-C^{2v}_{\rho^{}})
  (C^{2v}_{\rho^{'}}-C^{2v}_{\rho^{}})}-\label{d32}\\
\nonumber &-&(f^{(2)}_{\rho^{'''} NN}/f_{\rho^{'''}})
 \frac{(C^{2v}_{\rho^{'''}}-C^{2v}_{\rho^{''}})
       (C^{2v}_{\rho^{'''}}-C^{2v}_{\rho^{'}})}
       {(C^{2v}_{\rho^{''}}-C^{2v}_{\rho^{}})
       (C^{2v}_{\rho^{'}}-C^{2v}_{\rho^{}})}-\\
\nonumber &-&
  (f^{(2)}_{\rho^{''''} NN}/f_{\rho^{''''}})
  \frac{(C^{2v}_{\rho^{''''}}-C^{2v}_{\rho^{''}})
       (C^{2v}_{\rho^{''''}}-C^{2v}_{\rho^{'}})}
       {(C^{2v}_{\rho^{''}}-C^{2v}_{\rho^{}})
       (C^{2v}_{\rho^{'}}-C^{2v}_{\rho^{}})}\\
\nonumber (f^{(2)}_{\rho^{'} NN}/f_{\rho^{'}})&=&-\frac{1}{2}
 (\mu_p-\mu_n)\frac{C^{2v}_{\rho^{''}}C^{2v}_{\rho^{}}}
 {(C^{2v}_{\rho^{''}}-C^{2v}_{\rho^{'}})
  (C^{2v}_{\rho^{'}}-C^{2v}_{\rho^{}})}+\\
\nonumber &+&(f^{(2)}_{\rho^{'''} NN}/f_{\rho^{'''}})
 \frac{(C^{2v}_{\rho^{'''}}-C^{2v}_{\rho^{''}})
       (C^{2v}_{\rho^{'''}}-C^{2v}_{\rho^{}})}
       {(C^{2v}_{\rho^{''}}-C^{2v}_{\rho^{'}})
       (C^{2v}_{\rho^{'}}-C^{2v}_{\rho^{}})}+\\
\nonumber &+&
  (f^{(2)}_{\rho^{''''} NN}/f_{\rho^{''''}})
  \frac{(C^{2v}_{\rho^{'''}}-C^{2v}_{\rho^{''}})
       (C^{2v}_{\rho^{''''}}-C^{2v}_{\rho^{}})}
       {(C^{2v}_{\rho^{''}}-C^{2v}_{\rho^{'}})
       (C^{2v}_{\rho^{'}}-C^{2v}_{\rho^{}})}\\
\nonumber (f^{(2)}_{\rho^{''} NN}/f_{\rho^{''}})&=&\frac{1}{2}
 (\mu_p-\mu_n)\frac{C^{2v}_{\rho^{'}}C^{2v}_{\rho^{}}}
 {(C^{2v}_{\rho^{''}}-C^{2v}_{\rho^{'}})
  (C^{2v}_{\rho^{''}}-C^{2v}_{\rho^{}})}-\\
\nonumber &-&(f^{(2)}_{\rho^{'''} NN}/f_{\rho^{'''}})
 \frac{(C^{2v}_{\rho^{'''}}-C^{2v}_{\rho^{'}})
       (C^{2v}_{\rho^{'''}}-C^{2v}_{\rho^{}})}
       {(C^{2v}_{\rho^{''}}-C^{2v}_{\rho^{'}})
       (C^{2v}_{\rho^{''}}-C^{2v}_{\rho^{}})}-\\
\nonumber &-&
  (f^{(2)}_{\rho^{''''} NN}/f_{\rho^{''''}})
  \frac{(C^{2v}_{\rho^{''''}}-C^{2v}_{\rho^{'}})
       (C^{2v}_{\rho^{''''}}-C^{2v}_{\rho^{}})}
       {(C^{2v}_{\rho^{''}}-C^{2v}_{\rho^{'}})
       (C^{2v}_{\rho^{''}}-C^{2v}_{\rho^{}})}
\end{eqnarray}
the following coupling constant ratio numerical values are predicted
\begin{eqnarray}
\nonumber &(f^{(1)}_{\omega^{'}NN}/f_{\omega^{'}})=0.5045
&(f^{(1)}_{\rho^{'}NN}/f_{\rho^{'}})=0.7647\\
\nonumber &(f^{(1)}_{\omega^{''}NN}/f_{\omega^{''}})=0.1482
&(f^{(1)}_{\rho^{''}NN}/f_{\rho^{''}})=-0.6199\\
&(f^{(2)}_{\omega^{}NN}/f_{\omega^{}})=0.1712
&(f^{(2)}_{\rho^{}NN}/f_{\rho^{}})=3.0530\label{d33}\\
\nonumber &(f^{(2)}_{\omega^{'}NN}/f_{\omega^{'}})=-0.02455
&(f^{(2)}_{\rho^{'}NN}/f_{\rho^{'}})=-1.6790\\
\nonumber &(f^{(2)}_{\omega^{''}NN}/f_{\omega^{''}})=-0.05992
&(f^{(2)}_{\rho^{''}NN}/f_{\rho^{''}})=1.0040.
\end{eqnarray}

   The universal vector meson coupling constants $f_s$ and $f_v$ are determined
from the leptonic decay widths by the relation
\begin{equation}
\frac{f_v^2}{4\pi}=\frac{\alpha^2}{3}\frac{m_v}{\Gamma(\nu\to e^+e^-)}.
\label{d34}\end{equation}
Then, numerical values
\begin{equation}
f_\rho=5.0320\pm 0.1089; \; f_\omega= 17.0499\pm 0.2990;\; f_\phi=-12.8832\pm 0.0824
\label{d35}\end{equation}
are found from the corresponding world averaged lepton widths \ct{33}
and the universal $\omega^{'}-$, $\omega^{''}-$ and $\rho^{'}-$, $\rho^{''}-$
meson coupling constants
\begin{eqnarray}
f_{\omega^{'}}=47.6022\pm 7.5026;\;f_{\omega^{''}}=48.3778\pm 7.5026
\label{d36}\end{eqnarray}
and
\begin{eqnarray}
f_{\rho^{'}}=13.6491\pm 0.9521;\;f_{\rho^{''}}=22.4020\pm 2.2728
\label{d37}\end{eqnarray}
have been determined from the leptonic widths estimated by Donnachie and Clegg \ct{37}.

As a result, the following numerical values of the corresponding coupling
constants are predicted
\begin{eqnarray}
\nonumber &f^{(1)}_{\omega^{}NN}=18.9527; &f^{(1)}_{\rho^{}NN}=1.9335; \\
\nonumber &f^{(1)}_{\phi^{}NN}= 12.0956; &f^{(1)}_{\rho^{'}NN}=10.4375; \\
\nonumber &f^{(1)}_{\omega^{'}NN}=24.0153; &f^{(1)}_{\rho^{''}NN}=-13.8870; \\
\nonumber &f^{(1)}_{\omega^{''}NN}=7.1696; & \\
          &\label{d38}\\
\nonumber &f^{(2)}_{\omega^{}NN}=2.9189; &f^{(2)}_{\rho^{}NN}=15.3627; \\
\nonumber &f^{(2)}_{\phi^{}NN}= 3.4251; &f^{(2)}_{\rho^{'}NN}=-22.9168; \\
\nonumber &f^{(2)}_{\omega^{'}NN}=-1.1686; &f^{(2)}_{\rho^{''}NN}= 22.4916;\\
\nonumber &f^{(2)}_{\omega^{''}NN}=-2.8988. &
\end{eqnarray}

\begin{table}[thb]
\caption{\bf Coupling constants of the isoscalar vector mesons to
nucleons}
\begin{center}
%\begin{minipage}{12cm}
\begin{tabular}{|c|c|c|c|c|} \hline
 &$f_{\omega NN}^{(1)2}/4\pi$&$f_{\phi NN}^{(1)2}/4\pi$ & $f_{\omega' NN}^{(1)2}/4\pi$ & $f_{\omega'' NN}^{(1)2}/4\pi$ \\
 \hline \hline
our results&28.58&11.64&45.89&4.09  \\ \hline
Ref.[27]&34.6&6.7&--&--       \\ \hline Ref. [30]&24.0&5.1&--&--
\\ \hline
\end{tabular}
\begin{tabular}{|c|c|c|c|c|c|} \hline
 & $f_{\omega NN}^{(2)2}/4\pi$ & $f_{\phi NN}^{(2)2}/4\pi$ & $f_{\omega' NN}^{(2)2}/4\pi$ & $f_{\omega'' NN}^{(2)2}/4\pi$\\
 \hline \hline
our results&0.67&0.93&0.11&0.67  \\ \hline Ref.[27]&0.9&0.3&--&--
\\ \hline Ref. [30]&--&0.2&--&--    \\ \hline
\end{tabular}
%\end{minipage}
\end{center}
\end{table}

\begin{table}[thb]
\caption{\bf Coupling constants of the isovector vector mesons to
nucleons}
\begin{center}
%\begin{minipage}{12cm}
\begin{tabular}{|c|c|c|c|} \hline
 &$f_{\rho NN}^{(1)2}/4\pi$&$f_{\rho' NN}^{(1)2}/4\pi$ & $f_{\rho'' NN}^{(1)2}/4\pi$  \\
 \hline \hline
our results&0.30&8.67&15.35 \\ \hline Ref.[27]&--&40.27&793.53
\\ \hline Ref. [30]&0.55&--&--    \\ \hline
\end{tabular}
\begin{tabular}{|c|c|c|c|c|c|} \hline
 & $f_{\rho NN}^{(2)2}/4\pi$ & $f_{\rho' NN}^{(2)2}/4\pi$ & $f_{\rho'' NN}^{(2)2}/4\pi$  \\
 \hline \hline
our results&18.78&41.79&40.26  \\ \hline Ref.[27]&--&143.97&304.07
\\ \hline Ref. [30]&24.0&11.5&--    \\ \hline
\end{tabular}
%\end{minipage}
%\caption{Coupling constants of the isovector vector mesons to nucleons}
\end{center}
\end{table}

Their squares divided by $4\pi$ are reviewed in Table 2 and Table 3,
where for  comparison also values obtained by other authors \ct{27,30} are presented.

   One can immediately notice  large value of the $f^{(1,2)}_{\phi NN}$ coupling
constants which may indicate  violation of the OZI rule \ct{38}.

   Using the numerical values (\ref{d38}) one can predict the
$\omega-\phi$ mixing angle employing the relation
\begin{equation}
\frac{\sqrt{3}}{\cos\vartheta}\frac{f^{(1)}_{\rho NN}}{f^{(1)}_{\omega NN}}
-\tan\vartheta=\frac{f^{(1)}_{\phi NN}}{f^{(1)}_{\omega NN}}.\label{d39}
\end{equation}
It takes the value $\vartheta=0.7175$  which is very close to the ideal mixing.

\begin{figure}[thb]
\begin{center}
\hspace{2cm}\psfig{figure=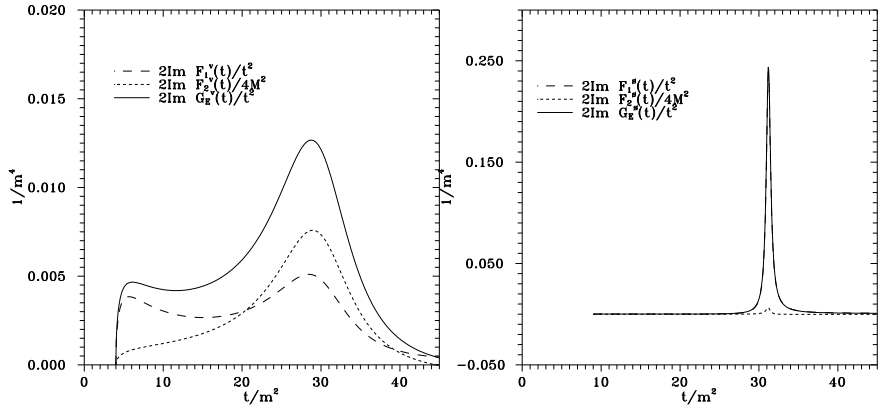,width=12cm}
\end{center} \caption{Predicted behaviours of the isovector and
isoscalar spectral functions by the ten-resonance unitary and
analytic model of the nucleon e.m. structure}
%\label{bari}
\end{figure}

   Nevertheless, the most important predictions of the unitary and analytic
model of the nucleon e.m. structure are the isovector spectral
function behaviours (see Fig. 6) to be consistent with the
predictions of H\H{o}hler and Pietarinen \ct{31} and Mergel,
Mei\ss ner and Drechsel \ct{27}, which have been carried out on
the basis of the Frazer and Fulco \ct{39} unitarity relation by
using the pion e.m. ff $F_\pi(t)$ and the $P$-wave $\pi\pi\to
N\bar N$ partial wave amplitudes obtained by an analytic
continuation of  experimental information on $\pi N$-scattering
into the unphysical region.

   The method of our prediction of the latter consists in the following. The ten-resonance
unitary and analytic model of the nucleon e.m. structure
constructed in this paper contains an explicit two-pion continuum
contribution given by the unitary cut starting with $t=4m_\pi^2$
from where just the isovector spectral functions start to be
different from zero. Then, despite  the fact that the unstable
$\rho$-meson is taken into account as a pole shifted from the real
axis into the complex plane on the second Riemann sheet of the
four-sheeted Riemann surface, the model predicts the strong
enhancement on the left wing of the $\rho$(770) resonance in the
isovector spectral functions automatically. Just agreement of our
predictions with those obtained by means of the Frazer and Fulco
unitarity relation convinces us that our model constructed in this
paper is really unitary.

   Another result of the presented model is the prediction of the isoscalar nucleon
spectral function behaviours (see Fig. 6),  as the model contains
an explicit three-pion continuum contribution given by the unitary
cut starting with $t=9m_\pi^2$  from where just the isoscalar
spectral functions start to be different from zero.

\section{Conclusions}

   We have constructed the unitary and analytic ten-resonance (5 isoscalars and 5
isovectors) model of the nucleon e.m. structure which represents a harmonic
unification of all known nucleon ff properties, like analyticity, reality
condition, experimental fact of creation of vector-meson resonances in
electron-positron annihilation processes, normalization and the asymptotic behaviour as
predicted for nucleon e.m. ff's by the quark model of hadrons. It depends
only on parameters with clear physical meaning. They are four effective
square-root branch points  representing contribution of all other higher
thresholds given by the unitarity condition, the mass and width of the
hypothetical fourth excited state of the $\rho$(770)-meson and coupling constants
of some resonances under consideration. They all are numerically evaluated
by  analyzing  all existing space-like and time-like nucleon ff data.

   We would like to note that by means of the model presented in this paper
all existing nucleon ff data, including
 FENICE neutron time-like data and FERMILAB proton eight points
at higher energies,  are reasonably described.  In this effect,  existence of the
$\rho^{''''}$(2500) resonance with the parameters $m_{\rho^{''''}}=2455\;MeV$ and
$\Gamma_{\rho^{''''}}=728\;MeV$ plays a crucial role. So, there is challenge to experimental
physicists to confirm  existence of this resonance also in other processes than
$e^+e^-\to N\bar N$.

   The unitary and analytic ten-resonance nucleon ff model gives several
reasonable predictions. However, the most important among them are isoscalar
and isovector spectral function behaviours  which coincide also with
the predictions obtained in the framework of heavy baryon chiral perturbation
theory \ct{40}.

 The work was  supported in part by the Slovak Grant Agency for Sciences,
Grant No. 2/1111/21(S.D.) and Grant No. 1/7068/21(A.Z.D).

\section*{Appendix A}
\renewcommand{\theequation}{A.\arabic{equation}}
\setcounter{equation}{0}

   The isoscalar and isovector parts of the Dirac and Pauli ff's are saturated by the isoscalar
and isovector vector-mesons as follows:

\begin{eqnarray}
\nonumber F_1^s(t)=\sum_{\sa,\sd,\sbb,\sc,\se}\mmm s\fff 1{s};
& &F_1^v(t)=\sum_{\va,\vb,\vc,\vd,\ve}\mmm v\fff 1{v};\\
F_2^s(t)=\sum_{\sa,\sd,\sbb,\sc,\se}\mmm s\fff 2{s};
& &F_2^v(t)=\sum_{\va,\vb,\vc,\vd,\ve}\mmm v\fff 2{v},\label{a1}
\end{eqnarray}
where $m_s$ and $m_v$ are the isoscalar and isovector vector-meson
masses, $f^{(1)}_{sNN}$, $f^{(1)}_{vNN}$ and $f^{(2)}_{sNN}$,
$f^{(2)}_{vNN}$ are the vector and tensor vector-meson-nucleon coupling
constants and $f_s$, $f_v$ are the universal vector-meson coupling
constants to be determined in a vector-meson decay into two charged
leptons. The explicit requirement of normalizations (6) and asymptotic
behaviours (7) in (\ref{a1}) leads to four systems of algebraic equations \ct{35}

\begin{eqnarray}
\nonumber {\rm I}. & &\sum_{\sa,\sd,\sbb,\sc,\se}\fff 1{s}=\frac{1}{2}\\
\  & &\sum_{\sa,\sd,\sbb,\sc,\se}\fff 1{s}m_s^2=0\label{a2}\\
\nonumber\ & &\\
\nonumber {\rm II}. & &\sum_{\va,\vb,\vc,\vd,\ve}\fff 1{v}=\frac{1}{2}\\
\  & &\sum_{\va,\vb,\vc,\vd,\ve}\fff 1{v}m_v^2=0\label{a3}\\
\nonumber\ & &\\
\nonumber {\rm III}. & &\sum_{\sa,\sd,\sbb,\sc,\se}\fff 2{s}=\frac{1}{2}(\mu_p+\mu_n)\\
\   & &\sum_{\sa,\sd,\sbb,\sc,\se}\fff 2{s}m_s^2=0 \label{a4}\\
\   & &\sum_{\sa,\sd,\sbb,\sc,\se}\fff 2{s}m_s^4=0 \nonumber \\
\nonumber {\rm IV}. & &\sum_{\va,\vb,\vc,\vd,\ve}\fff 2{v}=\frac{1}{2}(\mu_p-\mu_n)\\
\   & &\sum_{\va,\vb,\vc,\vd,\ve}\fff 2{v}m_v^2=0 \label{a5}\\
\   & &\sum_{\va,\vb,\vc,\vd,\ve}\fff 2{v}m_v^4=0
\nonumber
\end{eqnarray}
for $\fff 1{s}$, $\fff 1{v}$, $\fff 2{s}$ and $\fff 2{v}$.
The solutions of (\ref{a2})-(\ref{a5}) can be chosen in the following form:
\begin{eqnarray}
\nonumber {\rm I.} \;\;\; \fff {1}{\sbb}&=&\frac{1}{2}\frac{\mcsc}{\mcsc-\mcsb}
  -\fff {1}{\sa}\zll{\mcsc}{\mcsa}{\mcsc}{\mcsb}-\\
&-&\fff {1}\sd\zll{\mcsc}{\mcsd}{\mcsc}{\mcsb}+
  \fff {1}\se\zll{\mcse}{\mcsc}{\mcsc}{\mcsb}\label{a6}\\
\nonumber
\fff {1}{\sc}&=&-\frac{1}{2}\frac{\mcsb}{\mcsc-\mcsb}
  +\fff {1}{\sa}\zll{\mcsb}{\mcsa}{\mcsc}{\mcsb}+\\
\nonumber
&+&\fff {1}\sd\zll{\mcsb}{\mcsd}{\mcsc}{\mcsb}
  -\fff {1}\se\zll{\mcse}{\mcsb}{\mcsc}{\mcsb}\\
\nonumber
& &\qquad\\
\nonumber
{\rm II.} \;\;\;\fff {1}{\vb}&=&\frac{1}{2}\frac{\mcvc}{\mcvc-\mcvb}
  -\fff {1}{\va}\zll{\mcvc}{\mcva}{\mcvc}{\mcvb}+\\
&+&\fff {1}\vd\zll{\mcvd}{\mcvc}{\mcvc}{\mcvb}
  +\fff {1}\ve\zll{\mcve}{\mcvc}{\mcvc}{\mcvb}\label{a7}\\
\nonumber
 \fff {1}{\vc}&=&-\frac{1}{2}\frac{\mcvb}{\mcvc-\mcvb}
  +\fff {1}{\va}\zll{\mcvb}{\mcva}{\mcvc}{\mcvb}-\\
\nonumber
&-&\fff {1}\vd\zll{\mcvd}{\mcvb}{\mcvc}{\mcvb}
  -\fff {1}\ve\zll{\mcve}{\mcvb}{\mcvc}{\mcvb}\\
\nonumber
& &\qquad\\
\nonumber
{\rm III.}\;\;\; \fff 2{\sa}&=&\frac{1}{2}(\mu_p+\mu_n)\zzl\mcsc\mcsb\mcsc\mcsa\mcsb\mcsa-\\
\nonumber
  &-&\fff 2{\sd}\zlll\mcsc\mcsd\mcsb\mcsd\mcsc\mcsa\mcsb\mcsa-\\
\nonumber
  &-&\fff 2{\se}\zlll\mcse\mcsc\mcse\mcsb\mcsc\mcsa\mcsb\mcsa\\
\nonumber
 \fff 2{\sbb}&=&-\frac{1}{2}(\mu_p+\mu_n)\zzl\mcsc\mcsa\mcsc\mcsb\mcsb\mcsa-\\
 &-&\fff 2{\sd}\zlll\mcsc\mcsd\mcsd\mcsa\mcsc\mcsb\mcsb\mcsa+\label{a8}\\
\nonumber
  &+&\fff 2{\se}\zlll\mcse\mcsc\mcse\mcsa\mcsc\mcsb\mcsb\mcsa\\
\nonumber
 \fff 2{\sc}&=&\frac{1}{2}(\mu_p+\mu_n)\zzl\mcsb\mcsa\mcsc\mcsb\mcsc\mcsa+\\
\nonumber
 &+&\fff 2{\sd}\zlll\mcsb\mcsd\mcsd\mcsa\mcsc\mcsb\mcsc\mcsa-\\
\nonumber
 &-&\fff 2{\se}\zlll\mcse\mcsb\mcse\mcsa\mcsc\mcsb\mcsc\mcsa\\
\nonumber
& &\qquad\\
\nonumber
{\rm IV.}\;\;\; \fff 2{\va}&=&\frac{1}{2}(\mu_p-\mu_n)\zzl\mcvc\mcvb\mcvc\mcva\mcvb\mcva-\\
\nonumber
   &-&\fff 2{\vd}\zlll\mcvd\mcvc\mcvd\mcvb\mcvc\mcva\mcvb\mcva-\\
\nonumber
   &-&\fff 2{\ve}\zlll\mcve\mcvc\mcve\mcvb\mcvc\mcva\mcvb\mcva\\
\nonumber
 \fff 2{\vb}&=&-\frac{1}{2}(\mu_p-\mu_n)\zzl\mcvc\mcva\mcvc\mcvb\mcvb\mcva+\\
   &+&\fff 2{\vd}\zlll\mcvd\mcvc\mcvd\mcva\mcvc\mcvb\mcvb\mcva+\label{a9}\\
\nonumber
   &+&\fff 2{\ve}\zlll\mcve\mcvc\mcve\mcva\mcvc\mcvb\mcvb\mcva\\
\nonumber
 \fff 2{\vc}&=&\frac{1}{2}(\mu_p-\mu_n)\zzl\mcvb\mcva\mcvc\mcvb\mcvc\mcva-\\
\nonumber
   &-&\fff 2{\vd}\zlll\mcvd\mcvb\mcvd\mcva\mcvc\mcvb\mcvc\mcva-\\
\nonumber
   &-&\fff 2{\ve}\zlll\mcve\mcvb\mcve\mcva\mcvc\mcvb\mcvc\mcva,
\end{eqnarray}
which transform the original parametrizations (\ref{a1}) of the
isoscalar and isovector Dirac and Pauli nucleon ff's just into the
normalized zero-width VMD expressions (\ref{d8})-(\ref{d11}) with
asymptotics (\ref{d7}).

\section*{Appendix B}
\renewcommand{\theequation}{B.\arabic{equation}}
\setcounter{equation}{0}

   Incorporation of the assumed analytic properties of the nucleon e.m. ff's
into the normalized zero-width VMD model  (\ref{d8})-(\ref{d11}) can be
achieved by   application of the nonlinear transformations (\ref{d12})
and  a subsequent installation of the nonzero values of vector meson widths.

   There are also other expressions utilized for the vector meson masses squared
\begin{eqnarray}
\nonumber\mc s=t_0^s-\frac{4(t_{in}^{1s}-t_0^s)}{[1/V_{s0}-V_{s0}]^2},
& &\mc s=t_0^s-\frac{4(t_{in}^{2s}-t_0^s)}{[1/U_{s0}-U_{s0}]^2},\\
\mc v=t_0^v-\frac{4(t_{in}^{1v}-t_0^v)}{[1/W_{v0}-W_{v0}]^2},
& &\mc v=t_0^v-\frac{4(t_{in}^{2v}-t_0^v)}{[1/X_{v0}-X_{v0}]^2},\label{b1}
\end{eqnarray}
and identities
\begin{eqnarray}
\nonumber 0=t_0^s-\frac{4(t_{in}^{1s}-t_0^s)}{[1/V_N-V_N]^2},
& &0=t_0^s-\frac{4(t_{in}^{2s}-t_0^s)}{[1/U_N-U_N]^2},\\
0=t_0^v-\frac{4(t_{in}^{1v}-t_0^v)}{[1/W_N-W_N]^2},
& &0=t_0^v-\frac{4(t_{in}^{2v}-t_0^v)}{[1/X_N-X_N]^2},\label{b2}
\end{eqnarray}
following from (\ref{d12})  where $V_{s0}$, $W_{v0}$, $U_{s0}$, $X_{v0}$ are
the zero-width (therefore, they have a subindex 0) VMD poles and $V_N$, $W_N$,
$U_N$, $X_N$ are the normalization points (corresponding to $t=0$) in the
$V$, $W$, $U$, $X$ planes, respectively.

   Really,  relations (\ref{d12}),  (\ref{b1}), (\ref{b2}) first transform every t-dependent term
and every constant term consisting of a ratio of mass differences in (\ref{d8})-(\ref{d11})
into a new form as follows. For instance, the term $\mc\sa/(\mc\sa-t)$
in (\ref{d8}) is transformed into the following form:
\begin{equation}
\mmm\omega=\frac{\mcsa-0}{\mcsa-t}=\left(\frac{1-V^2}{1-V_N^2}\right)^2\frac
 {(V_N-V_{\sam})(V_N+V_{\sam})(V_N-1/V_{\sam})(V_N+1/V_{\sam})}
 {(V-V_{\sam})(V+V_{\sam})(V-1/V_{\sam})(V+1/V_{\sam})}.\label{b3}
\end{equation}
The constant mass terms, e.g. $(\mcsb-\mcsa)/(\mcsc-\mcsb)$, also from (\ref{d8}),
becomes:
\begin{eqnarray}
& &\frac{\mcsb-\mcsa}{\mcsc-\mcsb}=
\nonumber\frac{(\mcsb-0)-(\mcsa-0)}{(\mcsc-0)-(\mcsb-0)}=\\
\nonumber & &=\left [\frac{(V_N-V_{\sbbm})(V_N+V_{\sbbm})(V_N-1/V_{\sbbm})(V_N+1/V_{\sbbm})}
  {(V_{\sbbm}-1/V_{\sbbm})^2}\right.-\\
& &\left.-\frac{(V_N-V_{\sam})(V_N+V_{\sam})(V_N-1/V_{\sam})(V_N+1/V_{\sam})}
  {(V_{\sam}-1/V_{\sam})^2}\right ]/\label{b4}\\
\nonumber & &\left[\frac{(V_N-V_{\scm})(V_N+V_{\scm})(V_N-1/V_{\scm})(V_N+1/V_{\scm})}
  {(V_{\scm}-1/V_{\scm})^2}\right.-\\
\nonumber & &\left.-\frac{(V_N-V_{\sbbm})(V_N+V_{\sbbm})(V_N-1/V_{\sbbm})(V_N+1/V_{\sbbm})}
  {(V_{\sbbm}-1/V_{\sbbm})^2}\right ]=\\
\nonumber & &=
\frac{C_{{\omega'}_0}^{1s}-C_{{\omega}_0}^{1s}}{C_{{\omega''}_0}^{1s}-C_{{\omega'}_0}^{1s}}.
\end{eqnarray}

   Then by utilization of the relations between complex and complex conjugate values of the
corresponding zero-width VMD pole positions in the $V$, $W$, $U$, $X$ planes
\begin{eqnarray}
\nonumber & &V_{\sam}=-V_{\sam}^*;\;V_{\sdm}=-V_{\sdm}^*;\;V_{\sbbm}=-V_{\sbbm}^*;\;
 V_{\scm}=1/V_{\scm}^*;\;V_{\sem}=1/V_{\sem}^*\\
& &W_{\vam}=-W_{\vam}^*;\;W_{\vbm}=-W_{\vbm}^*;\;W_{\vcm}=-W_{\vcm}^*;\;
 W_{\vdm}=1/W_{\vdm}^*;\;W_{\vem}=1/W_{\vem}^*\label{b5}\\
\nonumber & &U_{\sam}=-U_{\sam}^*;\;U_{\sdm}=-U_{\sdm}^*;\;U_{\sbbm}=-U_{\sbbm}^*;\;
 U_{\scm}=1/U_{\scm}^*;\;U_{\sem}=1/U_{\sem}^*\\
\nonumber & &X_{\vam}=-X_{\vam}^*;\;X_{\vbm}=-X_{\vbm}^*;\;X_{\vcm}=-X_{\vcm}^*;\;
 X_{\vdm}=1/X_{\vdm}^*;\;X_{\vem}=1/X_{\vem}^*
\end{eqnarray}
following from the fact that in a fitting procedure one finds
\begin{eqnarray}
\nonumber
& &\mcsa-\Gamma^2_{\sa}/4<t_{in}^{1s};\;
\mcsd-\Gamma^2_{\sd}/4<t_{in}^{1s};\;
\mcsb-\Gamma^2_{\sbb}/4<t_{in}^{1s};\;\\
\nonumber
& &\mcsc-\Gamma^2_{\sc}/4>t_{in}^{1s};\;
\mcse-\Gamma^2_{\se}/4>t_{in}^{1s};\\
\nonumber
& &\mcsa-\Gamma^2_{\sa}/4<t_{in}^{2s};\;
\mcsd-\Gamma^2_{\sd}/4<t_{in}^{2s};\;
\mcsb-\Gamma^2_{\sbb}/4<t_{in}^{2s};\;\\
& &\mcsc-\Gamma^2_{\sc}/4>t_{in}^{2s};\;
\mcse-\Gamma^2_{\se}/4>t_{in}^{2s};\label{b6}\\
\nonumber
& &\mcva-\Gamma^2_{\va}/4<t_{in}^{1v};\;
\mcvb-\Gamma^2_{\vb}/4<t_{in}^{1v};\;
\mcvc-\Gamma^2_{\vc}/4<t_{in}^{1v};\;\\
\nonumber
& &\mcvd-\Gamma^2_{\vd}/4>t_{in}^{1v};\;
\mcve-\Gamma^2_{\ve}/4>t_{in}^{1v};\\
\nonumber
& &\mcva-\Gamma^2_{\va}/4<t_{in}^{2v};\;
\mcvb-\Gamma^2_{\vb}/4<t_{in}^{2v};\;
\mcvc-\Gamma^2_{\vc}/4<t_{in}^{2v};\;\\
\nonumber
& &\mcvd-\Gamma^2_{\vd}/4>t_{in}^{2v};\;
\mcve-\Gamma^2_{\ve}/4>t_{in}^{2v};
\end{eqnarray}
and subsequent introduction of the non-zero values of vector-meson widths
$\Gamma\not= 0$ by the substitutions
\begin{equation}
\mc s\to(m_s-i\frac{\Gamma_s}{2})^2;\;\;\;\;\mc v\to(m_v-i\frac{\Gamma_v}{2})^2,
\label{b7}
\end{equation}
one comes to (\ref{d14})-(\ref{d17}).


\begin{thebibliography}{99}
\bibitem{1}
S. Dubni\v cka, Nuovo Cim. {\bf A100}, (1988) 1.
\bibitem{2}
R. C. Walker et al., Phys. Rev {\bf D49}, (1994) 5671.
\bibitem{3}
L. Andivahis et al., Phys. Rev. {\bf D50}, (1994) 5491.
\bibitem{4}
A. F. Sill et al., Phys. Rev. {\bf D48}, (1993) 29.
\bibitem{5}
A. Lung et al., Phys. Rev. Lett. {\bf 70}, (1993) 718.
\bibitem{6}
S. Rock et al., Phys. Rev. {\bf D46}, (1992) 24.
\bibitem{7}
P. Markowitz et al., Phys. Rev. {\bf C48}, (1993) 5.
\bibitem{8}
E. E. Bruins et al., Phys. Rev. Lett. {\bf 75}, (1995) 21.
\bibitem{9}
M. Mayerhoff et al., Phys. Lett. {\bf 327B}, (1994) 201.
\bibitem{10}
S. Platchkov et al., Nucl. Phys. {\bf A510}, (1990) 740.
\bibitem{11}
R. D. Ransome, Nucl. Phys. {\bf A666/667}, (2000) 106c.
\bibitem{12}
T. Eden et al., Phys. Rev. {\bf C50}, (1994) R1749.
\bibitem{13}
M. Ostrick et al., Phys. Rev. Lett. {\bf 83}, (1999) 276.
\bibitem{14}
C. Herberg et al., Eur. Phys. J. {\bf A5}, (1999) 131.
\bibitem{15}
J. Becker et al., Eur. Phys. J. {\bf A6}, (1999) 329.
\bibitem{16}
D. Rohe et al., Phys. Rev. Lett. {\bf 83}, (1999) 4257.
\bibitem{17}
I. Passchier et al., Phys. Rev. Lett. {\bf 82}, (1999) 4988.
\bibitem{18}
G. Bassompierre et al, Nuovo Cimento {\bf A73}, (1983) 347.
\bibitem{19}
B. Delcourt et al, Phys. Lett. {\bf 86B}, (1979) 395.
\bibitem{20}
D. Bisello et al, Nucl. Phys. {\bf B224}, (1893) 379.
\bibitem{21}
T.A. Armstrong et al, Phys. Rev. Lett. {\bf 70}, (1993) 1212.
\bibitem{22}
M. Ambrogiani et al, Phys. Rev {\bf D60}, (1999) 032002.
\bibitem{23}
D. Bisello et al, J. Phys. {\bf C48}, (1990) 23.
\bibitem{24}
G. Bardin et al, Nucl. Phys. {\bf B411}, (1994) 3.
\bibitem{25}
A. Antonelli et al, Phys. Lett. {\bf 334B}, (1994) 431.
\bibitem{26}
C. Voci, Nucl. Phys. {\bf A623}, (1997) 333c.
\bibitem{27}
P. Mergell, U.-G. Meissner and D. Drechsel, Nucl. Phys. {\bf A596}, (1996) 367.
\bibitem{28}
S. Furiuchi and D. Watanabe, Nuovo Cimento {\bf A110}, (1997) 577.
\bibitem{29}
H.-W. Hammer, Ulf-G. Meissner and D. Drechsel, Phys. Lett. {\bf 385B}, (1996) 343.
\bibitem{30}
G. H\H{o}hler et al, Nucl. Phys. {\bf B114}, (1976) 505.
\bibitem{31}
G. H\H{o}hler and E. Pietarinen, Phys. Lett. {\bf 53B}, (1975) 471.
\bibitem{32}
S. J. Brodsky and P. G. Lepage, Phys. Rev. {\bf D22}, (1980) 2157.
\bibitem{33}
D. E. Groom et al., European Physical Journal {\bf C15} (2000) 1.
\bibitem{34}
M. E. Biagini, S. Dubni\v cka, E. Etim and P. Kola\v r, Nuovo Cimento {\bf A104}, (1991) 363.
\bibitem{35}
C. Adamu\v s\v cin, A. Z. Dubni\v ckova, S. Dubni\v cka, R. Pekarik, P. Weisenpacher, hep-ph/0203175
\bibitem{36}
S. Kopecky et al, Phys. Rev. Lett. {\bf 74} (1995) 2427.
\bibitem{37}
A. Donnachie and A.B. Clegg, Z. Phys. {\bf C42} (1989) 663.
\bibitem{38}
H. Genz and G. H\H{o}hler, Phys. Lett. {\bf 61B} (1976) 389.
\bibitem{39}
W.R. Frazer and J.R. Fulco, Phys. Rev. {\bf 117} (1960) 1603,1609.
\bibitem{40}
V. Bernard, N. Kaiser and Ulf-G. Meissner, Nucl. Phys. {\bf A611} (1996) 429.
\end{thebibliography}
\end{document}